\documentclass[12pt]{article}
\usepackage[dvips]{graphicx}
\usepackage{amssymb}

\makeatletter
 \@addtoreset{equation}{section}
 
\makeatother

\newcommand{\maru}{\mbox{\tiny$\stackrel{\circ}{\scriptstyle\circ}$}}

\begin{document}


\begin{titlepage}

\renewcommand{\thefootnote}{\fnsymbol{footnote}}

\begin{flushright}
\begin{tabular}{l}
UTHEP-546\\
KEK-TH-1157\\
June 2007
\end{tabular}
\end{flushright}

\bigskip

\begin{center}
{\Large \bf
 D-brane States and Disk Amplitudes in
 
  $OSp$ Invariant Closed String Field Theory
}

\end{center}

\bigskip

\begin{center}
{\large
Yutaka Baba}${}^{a}$\footnote{e-mail:
        yutaka@het.ph.tsukuba.ac.jp},
{\large Nobuyuki Ishibashi}${}^{a}$\footnote{e-mail:
        ishibash@het.ph.tsukuba.ac.jp},
{\large Koichi Murakami}${}^{b}$\footnote{e-mail:
        koichi@post.kek.jp}
\end{center}

\begin{center}
${}^{a}${\it
Institute of Physics, University of Tsukuba,\\
Tsukuba, Ibaraki 305-8571, Japan}\\
\end{center}
\begin{center}
${}^{b}${\it High Energy Accelerator Research Organization (KEK),\\
Tsukuba, Ibaraki 305-0801, Japan}
\end{center}

\bigskip

\bigskip

\bigskip

\begin{abstract}
We construct solitonic states in the $OSp$ invariant string field 
theory, which are BRST invariant in the leading order of regularization 
parameter $\epsilon$. 
We calculate the disk amplitudes using these solitonic states and 
show that they describe D-branes and ghost D-branes.
\end{abstract}

\setcounter{footnote}{0}
\renewcommand{\thefootnote}{\arabic{footnote}}

\end{titlepage}

\section{Introduction}
D-branes have been playing a central role in string theory for
a number of years. 
They can be considered as soliton solutions of open string field theory. 
For example, in bosonic open string field theory,
Sen conjectured that D$p$-branes with $p<25$ are described
as unstable lump solutions~\cite{Sen:1999xm}
and this was tested in many papers starting with
\cite{Harvey:2000tv}\cite{deMelloKoch:2000ie}\cite{Moeller:2000jy}.
What we would like to study in this paper is how one can realize D-branes 
in closed string field theory. 

The closed string field theory that we consider here is the $OSp$ invariant 
string field theory~\cite{Siegel:1984ap} for bosonic strings.
 (See also~\cite{Neveu:1986cu}\cite{Uehara:1987qz}
\cite{Kugo:1987rq}\cite{Kawano:1992dp}.) 
The $OSp$ invariant string field theory is a covariantized version  of 
the light-cone gauge string field
theory~\cite{Kaku:1974zz}\cite{Mandelstam:1973jk}\cite{Cremmer:1974ej}. 
It is constructed so that the S-matrix elements of the light-cone theory 
are reproduced by using this formulation. 
However an extra time variable exists in the formulation,
and the action of this theory looks different from
that of the usual field theory. 
Therefore, the $OSp$ invariant string field theory should be considered 
as something like stochastic 
or Parisi-Sourlas type formulation of field 
theory~\cite{Parisi:1980ys}\cite{Parisi:1979ka}.
In our previous work~\cite{Baba:2007je}, 
treating the theory in such a manner, we constructed 
BRST invariant observables in the $OSp$ invariant string 
field theory, from which one can calculate the S-matrix elements 
of string theory.

In \cite{Baba:2007je}, only on-shell asymptotic states are considered 
and the observables are BRST invariant up to the nonlinear terms. 
In order to construct off-shell BRST invariant states, 
we should take the nonlinear terms into account. 
What we would like to do in this paper is to 
construct such states using boundary states for D-branes. 
We consider states which act as source terms in the string field
theory and have the effect of generating boundaries 
in the worldsheet. 
Imposing the condition that the states are BRST invariant in the leading 
order of regularization parameter $\epsilon$, 
we can fix the form of the states. 
These states can be considered as states in which there exist 
solitons corresponding to D-branes 
and ghost D-branes~\cite{Okuda:2006fb}. 
We can construct states with an arbitrary number of such solitons. 
We calculate the disk amplitudes using these states 
and show that the disk amplitudes of bosonic string theory 
in the presence of D-branes and ghost D-branes are reproduced.

In \cite{Baba:2006rs}, solitonic states 
corresponding to even number of D-branes or ghost D-branes 
were constructed by using the similarity between the string field theory
for noncritical strings~\cite{Ishibashi:1993pc} 
and the $OSp$ invariant string field theory. 
Although our construction in this paper is essentially the same as that 
in \cite{Baba:2006rs}, we get different results because of 
several reasons. 
Firstly, 
in \cite{Baba:2006rs}, we considered the solitonic operators in analogy 
to noncritical string theory~\cite{Fukuma:1996hj}\cite{Hanada:2004im}
and postulated the form of the vacuum amplitude
using this analogy.
We checked that the vacuum amplitude coincide with that for even number 
of D-branes in string theory. 
However,  
the vacuum amplitude is a constant which may be changed at will 
by changing the definition. 
In this paper, we rather calculate the disk amplitudes, 
which can be defined without such ambiguities. 
We show that the treatment of the solitonic states in \cite{Baba:2006rs} 
corresponds to considering even number of solitons. 
Secondly, 
in \cite{Baba:2006rs}, we defined the creation and the annihilation
operators corresponding to normalized boundary states
and performed calculations using these operators, 
which made the calculations rather indirect. 
In this paper, we do not introduce the artificial
``normalized boundary states'', 
and calculate the BRST transformation of the solitonic states directly. 
We find that a factor of $2$ was overlooked in the calculations 
of \cite{Baba:2006rs}.

The organization of this paper is as follows. 
In section \ref{sec:soliton}, we construct solitonic states 
in the $OSp$ invariant string field theory. 
Imposing the condition that the states are BRST invariant 
in the leading order of regularization parameter $\epsilon$, 
we can fix the form of the states. 
In section \ref{sec:disk}, we calculate disk amplitudes using our solitonic 
states and show that disk amplitudes 
in the presence of D-branes and ghost D-branes are reproduced 
including the normalizations.
Thus we identify the solitons with D-branes and ghost D-branes. 
Section \ref{sec:discussion} is devoted to discussions.
In appendix \ref{sec:conventions}, we summarize the formulation 
of the $OSp$ invariant string field theory. 
In appendices \ref{sec:tree} and \ref{sec:torus}, we present the details 
of calculations needed to show the BRST invariance
of the solitonic states. 

\section{BRST Invariant Solitonic States}\label{sec:soliton}

In this paper, the notations for the variables of the $OSp$ invariant
string field theory are the same as those used in \cite{Baba:2007je},
otherwise stated.
Those are summarized in appendix \ref{sec:conventions}.

Let us consider a D$p$-brane (or a ghost D$p$-brane)
that extends in the $X^{\mu}$ $(\mu=26,1,\ldots,p$)
directions and is located at $X^{i} = 0$ $(i=p+1,\ldots,25)$\footnote{
In this paper, we consider a flat noncompact space-time. 
We do not need any infrared regularization 
in the calculations here, in contrast to those in \cite{Baba:2006rs}.
}.
We  denote these directions by $X^{\mu}$ $(\mu \in \mathrm{N})$ and
$X^{i}$ $(i \in \mathrm{D})$, respectively.

In the $OSp$ invariant string field theory, 
the boundary state
$|B_{0}\rangle$ corresponding to the D$p$-brane
can be constructed as follows~\cite{Baba:2006rs}. 
The matter fields
$X^{\mu}(\tau,\sigma)$, $X^{i}(\tau,\sigma)$ and the ghost fields
$C(\tau,\sigma)$, $\bar{C}(\tau,\sigma)$
satisfy the following boundary conditions at $\tau=0$,
\begin{equation}
\partial_{\tau}X^{\mu} (0,\sigma) |B_{0}\rangle
 = 0~,\quad
 X^{i}(0, \sigma)|B_{0}\rangle = 0~,
 \quad
 C(0, \sigma)|B_{0}\rangle
 =\bar{C}(0,\sigma) |B_{0}\rangle =0~.
 \label{eq:boundarycond}
\end{equation}
It follows that the state $|B_{0} \rangle$
is expressed in terms of the oscillation modes as
\begin{eqnarray}
 |B_{0} \rangle
  = 
     \exp \left[ - \sum_{n=1}^{\infty} \frac{1}{n} \alpha^{N}_{-n} 
                  \tilde{\alpha}^{M}_{-n} D_{NM} \right]
                  |0\rangle
      (2\pi)^{p+1} \delta^{p+1}_{\mathrm{N}}(p)~,
\label{eq:boundarystate}
\end{eqnarray}
where 
$\delta^{p+1}_{\mathrm{N}} (p)$
denotes the delta function of the momentum in the directions
along the D$p$-brane
defined as
$\delta^{p+1}_{\mathrm{N}} (p)
  = \prod_{\mu \in \mathrm{N}} \delta(p_{\mu})$,
and $D_{NM}$ denotes
\begin{equation}
D_{NM}=D^{NM}=
 \bordermatrix{
         &    & & \mbox{\scriptsize $C$} & 
                  \mbox{\scriptsize $\bar{C}$} \cr
         & \delta_{\mu\nu} & & & \cr
         & & - \delta_{ij} & & \cr
         \mbox{\scriptsize $C$} & & & 0 & i \cr
         \mbox{\scriptsize $\bar{C}$} & & & -i & 0 }
       \qquad \mbox{with \  
         $\mu,\nu \in \mathrm{N}$~,
         $i,j \in \mathrm{D}$}~.
\label{eq:metricD}
\end{equation}

Since the norm of the boundary state $|B_{0}\rangle$ diverges,
we need to regularize it.
In order to do so, we introduce 
\begin{equation}
|B_{0}\rangle^{T}
  = e^{- \frac{T}{|\alpha|}(L_{0}+\tilde{L}_{0} -2)}
    |B_{0}\rangle
\end{equation}
for $T>0$,
and consider $|B_{0}\rangle^{\epsilon}$ with 
$0<\epsilon \ll 1$ as
a regularized version of $|B_{0}\rangle$.
Notice that the operator 
$e^{- \frac{T}{|\alpha|}(L_{0}+\tilde{L}_{0} -2)}$ 
commutes with the BRST operator $Q_{\mathrm{B}}$.

\subsection{States with one soliton}

Using the regularized boundary state $|B_{0}\rangle^{\epsilon}$,
let us construct a state in the following form
in the Hilbert space of the $OSp$ invariant string field theory,
\begin{equation}
|D \rangle \! \rangle
\equiv
\lambda \int d\zeta\, \bar{\mathcal{O}}_{D} (\zeta)
  |0\rangle\!\rangle~,
\label{eq:solstate}
\end{equation}
where 
\begin{equation}
\bar{\mathcal{O}}_{D}(\zeta) 
=
 \exp \left[ a
         \int^{0}_{-\infty} dr 
            \, \frac{e^{\zeta \alpha_{r}}}{\alpha_{r}}
         {}^{\epsilon}_{r} \langle B_{0} | \Phi \rangle_{r}
         +F(\zeta ) \right]~.
\label{eq:solitonicop}
\end{equation}
Here $\lambda$ and $a$ are constants,
$F(\zeta )$ is a function of $\zeta$ and
the limits of the zero-mode integration $dr$ in the exponent
of $\bar{\mathcal{O}}_D(\zeta)$
denotes the integration region of the string length $\alpha_{r}$.
Since the integration is over $-\infty <\alpha_r <0$, 
only the creation operators contribute to $\bar{\mathcal{O}}_D(\zeta)$.
Assuming that the integration over $\alpha_r$ 
is convergent with $\mathrm{Re}\,\zeta >0$
sufficiently large, we define $\bar{\mathcal{O}}_{D}(\zeta)$
by analytic continuation. 

Expanding the exponential in terms of the string field, 
it is easy to see that the state $|D\rangle \! \rangle$ 
has the effect of generating boundaries in the worldsheet, 
with a weight which depends on $a$ and $F(\zeta )$. 
Let us impose the condition that the state $|D\rangle \! \rangle$
is BRST invariant in the leading order of $\epsilon$. 
As we will see, we can determine $a$ and $F(\zeta )$ from this condition.

\subsubsection*{BRST transformation}

In order to evaluate $\delta_{\mathrm{B}} |D \rangle \! \rangle$,
we should calculate the BRST transformation of the operator in
the exponent of $\bar{\mathcal{O}}_D (\zeta)$:
\begin{eqnarray}
 \delta_{\mathrm{B}} \int^{0}_{-\infty} dr\,
 \frac{e^{\zeta \alpha_{r}}}{\alpha_{r}} 
  {}^{\epsilon}_{r} \langle B_{0} | \Phi \rangle_{r}
 &=& \int^{0}_{-\infty} dr \,\frac{ e^{\zeta \alpha_{r}}}{\alpha_{r}}
    {}^{\epsilon}_{r} \langle B_{0} |
     Q^{(r)}_{\mathrm{B}} |\Phi\rangle_{r}
     \nonumber\\
 && 
  {} + g \int^{\infty}_{0} d3 \,
    \frac{e^{- \zeta \alpha_{3}}}{\alpha_{3}}
     \int d1d2
     \, \langle V_{3}(1,2,3)|
       \Phi\rangle_{1} |\Phi\rangle_{2}
      |B_{0}\rangle_{3}^\epsilon~.
  \label{eq:brst-tr-exp1}
\end{eqnarray}

By using
\begin{equation}
Q_{\mathrm{B}} \left(\frac{1}{\alpha} |B_{0}\rangle^{\epsilon} \right)
 = 0~,
\end{equation}
one can recast the first term on the right hand side
of eq.(\ref{eq:brst-tr-exp1}) into
\begin{equation}
\int^{0}_{-\infty} dr \,\frac{ e^{\zeta \alpha_{r}}}{\alpha_{r}}
    {}^{\epsilon}_{r} \langle B_{0} |
     Q^{(r)}_{\mathrm{B}} |\Phi\rangle_{r}
= \zeta \int^{0}_{-\infty} dr \, \frac{e^{\zeta \alpha_{r}}}{\alpha_{r}}
  {}^{\epsilon}_{r} \langle B_{0} | i \pi^{(r)}_{0}
     | \bar{\psi} \rangle_{r}~.
\label{eq:brst-tr-exp-g0}
\end{equation}
Let us here introduce shorthand notations 
\begin{eqnarray}
\bar{\phi}(\zeta )
 & \equiv &
\int^{0}_{-\infty} dr\,
 \frac{e^{\zeta \alpha_{r}}}{\alpha_{r}} 
  {}^{\epsilon}_{r} \langle B_{0} | \Phi \rangle_{r}~,
\nonumber
\\
\bar{\chi}(\zeta )
&\equiv&
\int^{0}_{-\infty} dr \, \frac{e^{\zeta \alpha_{r}}}{\alpha_{r}}
  {}^{\epsilon}_{r} \langle B_{0} | i \pi^{(r)}_{0}
     | \bar{\psi} \rangle_{r}~,
\end{eqnarray}
in terms of which
eq.(\ref{eq:solitonicop}) can be expressed as
\begin{equation}
\bar{\mathcal{O}}_{D}(\zeta) 
=
\exp
\left(
a\bar{\phi}(\zeta )+F(\zeta )
\right)~,
\end{equation}
and 
eq.(\ref{eq:brst-tr-exp-g0}) can be written as
\begin{equation}
\int^{0}_{-\infty} dr \,\frac{ e^{\zeta \alpha_{r}}}{\alpha_{r}}
    {}^{\epsilon}_{r} \langle B_{0} |
     Q^{(r)}_{\mathrm{B}} |\Phi\rangle_{r}
     =\zeta \bar{\chi}(\zeta )~.
\end{equation}
Notice that $\bar{\phi}$ and $\bar{\chi}$ are made only from the 
creation modes and commute with each other. 

For the second term on the right hand side of eq.(\ref{eq:brst-tr-exp1}), 
we decompose $|\Phi\rangle$ into the creation and annihilation parts
as $|\Phi\rangle =|\psi\rangle +| \bar{\psi} \rangle$, 
and obtain
\begin{eqnarray}
\lefteqn{
g \int^{\infty}_{0} d3 \,
    \frac{e^{- \zeta \alpha_{3}}}{\alpha_{3}}
     \int d1d2
     \, \langle V_{3}(1,2,3)|
       \Phi\rangle_{1} |\Phi\rangle_{2}
      |B_{0}\rangle_{3}^\epsilon
} \nonumber\\
&& = g \int^{\infty}_{0} d3\, 
       \frac{e^{-\zeta \alpha_3}}{\alpha_{3}}
       \left[
       \int^{0}_{-\infty} d1 
       \int^{\infty}_{0} d2\,
       \langle V_{3}(1,2,3)| \bar{\psi} \rangle_{1}
       |\psi\rangle_{2} |B_{0}\rangle_{3}^{\epsilon}
    \right.  \nonumber\\
&& \hspace{8em}
  {} + \int^{\infty}_{0} d1 \int^{0}_{-\infty} d2
      \,
       \langle V_{3}(1,2,3)| \psi \rangle_1
       |\bar{\psi} \rangle_2 |B_{0}\rangle_3^{\epsilon}
    \nonumber\\
&& \hspace{8em} \left.
  {} + \int^{0}_{-\infty} d1 \int^{0}_{-\infty} d2 \,
      \langle V_{3}(1,2,3)| \bar{\psi} \rangle_1
       |\bar{\psi}  \rangle_2 |B_{0}\rangle_3^{\epsilon}
       \right]~.
\label{eq:brst-tr-exp-g1}
\end{eqnarray}
It follows from the relation
$ \langle V_{3}(1,2,3)|
  =  \langle V_{3}(2,1,3)|$
that the first and the second terms on the right hand side
of this equation are equal to each other.

\begin{figure}[htbp]
\begin{minipage}{18.5em}
\begin{center}
	\includegraphics[width=16.5em]{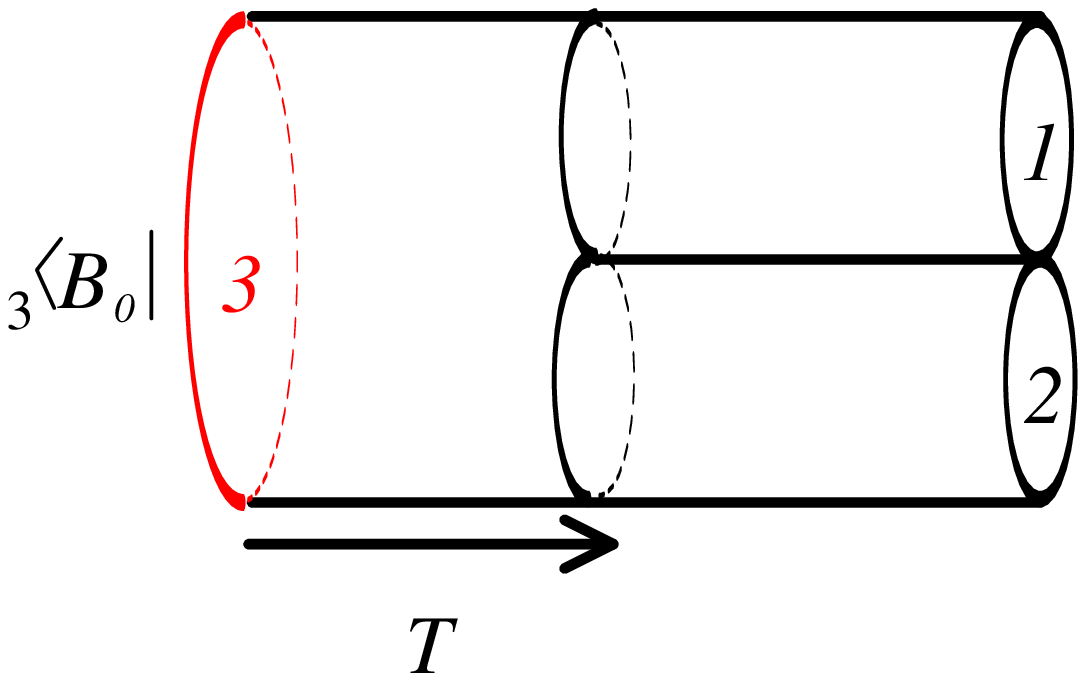}
	\caption{The string diagram corresponding to
	         the vertex $\langle V_{2}(1,2);T|$.}
	\label{fig:uhp1}
\end{center}
\end{minipage}
\hfill
\begin{minipage}{18.5em}
\begin{center}
	\includegraphics[width=16.5em]{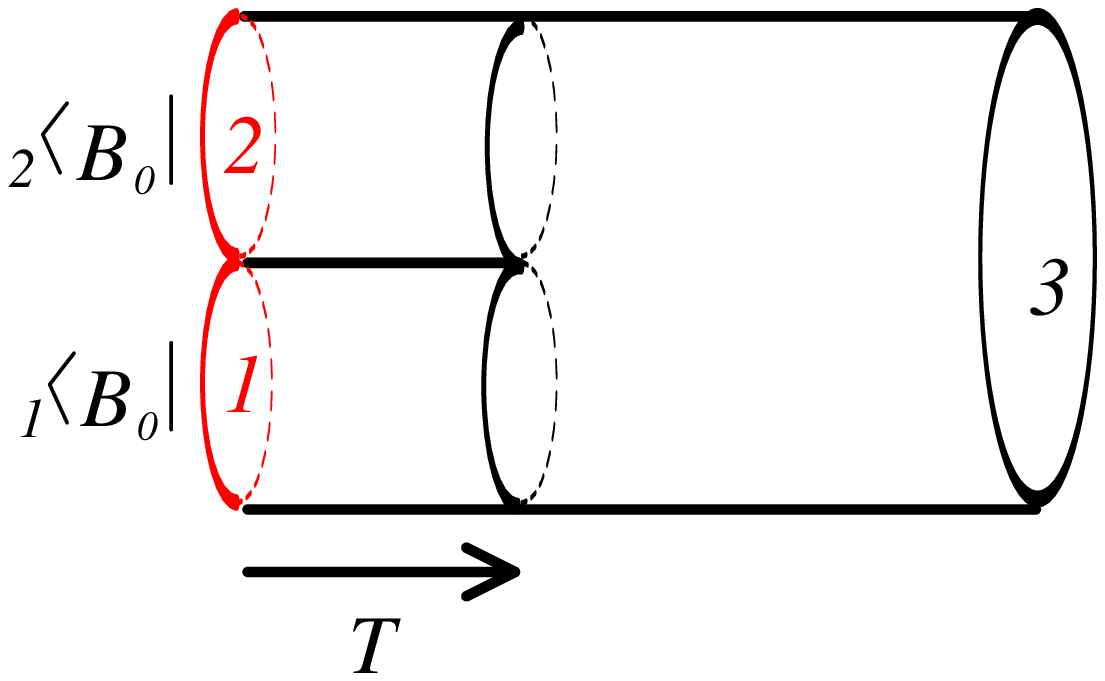}
	\caption{The string diagram corresponding to
	         the vertex $\langle V_{1}(3);T|$.}
	\label{fig:annulus1}
\end{center}
\end{minipage}
\end{figure}

In this form, it is straightforward to calculate the 
BRST transformation of $|D \rangle \! \rangle$.
Using the commutation relation (\ref{eq:ccr}),
we have
\begin{eqnarray}
\delta_{\mathrm{B}} |D \rangle \! \rangle 
&=&
\lambda \int d\zeta \,
\exp \left(a\bar{\phi}(\zeta )+F(\zeta )\right)
\nonumber\\
&&
\ \times
\left[ {}
a\zeta \bar{\chi}(\zeta )
{}+g a^2 \int^{0}_{-\infty} d1
         \int^{\infty}_{0} d2
         \int^{\infty}_{0} d3 
         \frac{e^{\zeta \alpha_{1}}}{\alpha_2 \alpha_3}
         \langle V_3(1,2,3)| \bar{\psi}\rangle_1
          |B_{0}\rangle^{\epsilon}_2|B_{0}\rangle^{\epsilon}_3
\right.  \nonumber\\
&&
\ \quad
 \left.
 {}+ga
   \int^{0}_{-\infty}d1
   \int^{0}_{-\infty} d2 
   \int^{\infty}_{0} d3 
   \frac{e^{\zeta (\alpha_1 + \alpha_2)}}{\alpha_3}
   \langle V_3(1,2,3)|\bar{\psi}\rangle_1
          |\bar{\psi}\rangle_2 |B_{0}\rangle^{\epsilon}_3
\right] 
|0\rangle\!\rangle~.~~~~
\label{eq:brsttr-D1}
\end{eqnarray}
This tells us that in order to evaluate the BRST transformation
of the state $|D\rangle \! \rangle$ we need to obtain
\begin{equation}
\langle V_2 (1,2);T| \equiv
\int d'3 \, 
  \langle V_3 (1,2,3)|B_{0}\rangle^{T}_{3}
\label{eq:treevertex}
\end{equation}
and
\begin{equation}
\langle V_1 (3); T| \equiv
\int d'1 d'2 \, 
\langle V_3 (1,2,3)|B_{0}\rangle^{T}_{1}
  |B_{0}\rangle^{T}_{2} 
\label{eq:loopvertex}
\end{equation}
for $T=\epsilon$.
Here $\alpha_{1}\alpha_{2} > 0$ in both cases. 
The integration measure $d'r$ is defined in eq.(\ref{eq:dprimer}).
These vertices respectively correspond to the string diagrams
depicted in Figs.\ref{fig:uhp1} and \ref{fig:annulus1}.

By using these vertices, 
eq.(\ref{eq:brsttr-D1}) can be rewritten as
\begin{eqnarray}
\delta_{\mathrm{B}} |D \rangle \! \rangle 
&=&
\lambda \int d\zeta 
\exp \left(a\bar{\phi}(\zeta )+F(\zeta )\right)
\nonumber\\
&&
\quad
\times
\left[ {}
a\zeta \bar{\chi}(\zeta )
{}+\frac{ga^2}{4} 
         \int^{\infty}_{0} d\alpha_1
         \int^{\infty}_{0} d\alpha_2
         \int^{0}_{-\infty} d3
         e^{\zeta \alpha_{3}}
         \langle V_1(3);\epsilon | \bar{\psi}\rangle_3
  \right.
  \nonumber\\
&&
\quad\quad
\left.
 {}+\frac{ga}{2}
   \int^{0}_{-\infty}d1
   \int^{0}_{-\infty} d2 
   \int^{\infty}_{0} d\alpha_3 
   e^{\zeta ( \alpha_1 + \alpha_{2}) }
   \langle V_2(1,2);\epsilon |\bar{\psi}\rangle_1
          |\bar{\psi}\rangle_2 
\right] 
|0\rangle\!\rangle~.  
\label{eq:brsttr-D2}
\end{eqnarray}
In appendices \ref{sec:tree} and \ref{sec:torus}, 
these vertices are computed in the leading order of $\epsilon$.
The results are
\begin{eqnarray}
\langle V_{2}(1,2);\epsilon| 
&\sim&
 2\delta (\alpha_{1}+\alpha_{2}+\alpha_{3})
 \times C_2 \times
 {}^{\epsilon}_{1}\langle B_{0}|
 \;  {}^{\epsilon}_{2}\langle B_{0}|
 \left( \frac{i}{\alpha_{1}} \pi^{(1)}_{0}
                  + \frac{i}{\alpha_{2}} \pi^{(2)}_{0} \right)
  \mathcal{P}_{12}~,
\\
\langle V_{1}(3);\epsilon |
 &\sim&
  - 2\delta (\alpha_{1}+\alpha_{2}+\alpha_{3})
  \times C_1 \times
      {}^{\epsilon}_{3} \langle B_{0}|
      \frac{2i}{\alpha_{3}}\pi^{(3)}_{0} \mathcal{P}_{3}~,
\end{eqnarray}
where
\begin{equation}
C_2
\equiv
\frac{1}{(16 \pi)^{\frac{p+1}{2}}}
 \frac{4}{\epsilon^{2} (- \ln \epsilon)^{\frac{p+1}{2}}}~,
\quad
C_1
\equiv
\frac{(4\pi^{3})^{\frac{p+1}{2}} }{(2\pi)^{25}}
      \frac{4}{\epsilon^{2}(-\ln \epsilon)^{\frac{p+1}{2}}}~.
\end{equation}
These are the idempotency equations~\cite{Kishimoto:2003ru} 
satisfied by the boundary states in the $OSp$ invariant string field theory. 
Substituting these into eq.(\ref{eq:brsttr-D2}), we obtain
\begin{equation}
\delta_{\mathrm{B}}|D\rangle\!\rangle
=
\lambda \int d\zeta \,
\Bigl[ a\zeta \bar{\chi}(\zeta )
       +ga^2 C_1\partial_\zeta \bar{\chi} (\zeta)
       + 2ga C_2\bar{\chi}(\zeta) \partial_\zeta \bar{\phi}(\zeta)
\Bigr]
\,
e^{a\bar{\phi}(\zeta )+F(\zeta )}
\, 
  |0 \rangle \! \rangle~.
\label{eq:deltaD}
\end{equation}
Here we have used the following identity
\begin{equation}
\int_0^\infty dl_1\int_0^\infty dl_2e^{-\zeta_1l_1-\zeta_2l_2}
f(l_1+l_2)
=
-\frac{\tilde{f}(\zeta_1 )-\tilde{f}(\zeta_2 )}{\zeta_1-\zeta_2}~,
\end{equation}
where
\begin{equation}
\tilde{f}(\zeta )
\equiv
\int_0^\infty dle^{-\zeta l}f(l)~.
\end{equation}

Now, in order to make $|D\rangle\!\rangle$ 
BRST invariant, we choose $F(\zeta )$ to be of the form
\begin{equation}
F(\zeta )=b\zeta^2~.
\end{equation}
Then 
the right hand side of eq.(\ref{eq:deltaD}) becomes
\begin{equation}
\lambda
\int
d\zeta\;
\partial_\zeta 
\left[
\frac{a}{2b}\bar{\chi}(\zeta )
\, \exp \left( a\bar{\phi}(\zeta )+b\zeta^2 \right)
\right]\,
|0 \rangle \! \rangle~,
\label{eq:totalder}
\end{equation}
provided the constants $a,b$ satisfy
\begin{equation}
\frac{a}{2b}
=ga^2C_1~,
\quad
\frac{a^2}{2b}
= 2gaC_2~.
\label{eq:abC1}
\end{equation}
These equations have the solutions $(a,b)=\pm ( A , B )$, where
\begin{equation}
A =
 \frac{(2\pi )^{13}}{(8\pi^2)^{\frac{p+1}{2}}\sqrt{\pi}}~,
\quad
B
= 
\frac{(2\pi )^{13}\epsilon^2(-\ln \epsilon )^{\frac{p+1}{2}}}
{16\left(\frac{\pi}{2}\right)^{\frac{p+1}{2}}\sqrt{\pi}g}~.
\label{eq:abpm}
\end{equation}
Therefore, by choosing $(a,b)$ as $\pm ( A , B )$ and 
taking the integration contour for $\zeta$ appropriately, 
we can obtain a state BRST invariant in the leading order of 
$\epsilon$. 
Let us define 
\begin{equation}
|D_{\pm}\rangle\!\rangle
\equiv \lambda_{\pm} \int d\zeta\, \bar{\mathcal{O}}_{D_{\pm}} (\zeta)
  |0\rangle\!\rangle~,
\label{eq:solitonicstatepm}
\end{equation}
with
\begin{equation}
\bar{\mathcal{O}}_{D_{\pm}}(\zeta) =
 \exp \left[ \pm \frac{(2 \pi)^{13}}
                      {(8 \pi^{2})^{\frac{p+1}{2}} \sqrt{\pi}}
         \int^{0}_{-\infty} dr 
            \, \frac{e^{\zeta \alpha_{r}}}{\alpha_{r}}
         {}^{\epsilon}_{r} \langle B_{0} | \bar{\psi}\rangle_{r}
         \pm \frac{(2 \pi)^{13}\epsilon^{2}
                               (- \ln \epsilon)^{\frac{p+1}{2}}}
                  {16 \left(\frac{\pi}{2}\right)^{\frac{p+1}{2}}
                          \sqrt{\pi} g}
             \zeta^{2} \right]~.
\label{eq:solitonicoppm}
\end{equation}

These states are considered as states
in which one D-brane or one ghost D-brane is excited. 
We will show that $|D_{\pm}\rangle\!\rangle $ generate the 
worldsheets with boundaries with the right weight 
and disk amplitudes are reproduced. 
In this paper, we take $g>0$. In this convention,
as we will see later, 
$|D_{+}\rangle\!\rangle $ corresponds to the D-brane and 
$|D_{-}\rangle\!\rangle $ corresponds to the ghost D-brane. 

One comment is in order. 
Here and in the following, we construct BRST invariant ket vectors 
in the second quantized Hilbert space. 
It is obvious that the hermitian conjugates of these ket vectors 
are also BRST invariant. 
Therefore the states $\langle\!\langle D_{\pm}|$ are BRST invariant. 

\subsection{States with $N$ solitons}
We can construct BRST invariant states with $N$ solitons in a similar way. 
Let us consider a state in the following form
\begin{equation}
|D_{N +} \rangle \! \rangle
\equiv
\lambda_{N+} \int \prod_{i=1}^Nd \zeta_i \, 
\bar{\mathcal{O}}_{D_{N+}} (\zeta_1,\cdots ,\zeta_N)
  |0\rangle\!\rangle~,
\label{eq:Nsolstate}
\end{equation}
where 
\begin{equation}
\bar{\mathcal{O}}_{D_{N +}}(\zeta_1,\cdots ,\zeta_N) 
=
 \exp \left[ \sum_{i=1}^N 
               \left(A \bar{\phi}(\zeta_i)
                     +B \zeta_i^2 \right)
            +F_N(\zeta_1,\cdots, \zeta_N)
      \right].
\label{eq:Nsolitonicop}
\end{equation}
Here the coefficients $A$ and $B$ are given in eq.(\ref{eq:abpm}),
and the function 
$F_N(\zeta_1,\cdots, \zeta_N)$ is to be determined. 

It is now straightforward to evaluate the BRST variation of this state:
\begin{eqnarray}
\delta_{\mathrm{B}}|D_{N+} \rangle\!\rangle
&=&
\lambda_{N+}  \int \prod_{i=1}^N d\zeta_i 
\exp
\left[  \sum_{i=1}^{N}
          \left( A  \bar{\phi}(\zeta_i )+ B  \zeta_i^2  \right)
+F_N(\zeta_1,\cdots ,\zeta_N)
\right]
\nonumber\\
&&
\qquad \quad
\times   \left[ \sum_{i=1}^{N}
  \Bigl(  A  \zeta_i \bar{\chi}(\zeta_i )
          +g A^2 C_1  \partial_{\zeta_i} \bar{\chi}(\zeta_i)
          + 2g A C_2  \bar{\chi}(\zeta_{i})
                       \partial_{\zeta_{i}} \bar{\phi}(\zeta_i)
  \Bigr)
\right.
\nonumber
\\
&&
\qquad\qquad \quad
\left.
+g A^2 C_1\sum_{i\neq j}
          \frac{\bar{\chi}(\zeta_i)-\bar{\chi}(\zeta_j)}{\zeta_i-\zeta_j}
          \right]
|0 \rangle \! \rangle~.
\label{eq:deltaDN}
\end{eqnarray}
Using eq.(\ref{eq:abC1}), one can easily deduce that 
the right hand side of eq.(\ref{eq:deltaDN}) can be recast into the form
\begin{equation}
\lambda_{N+} \int \prod_{i=1}^Nd\zeta_i
\sum_{j=1}^{N} \partial_{\zeta_j}
\left[
\frac{A}{2 B}
\bar{\chi}(\zeta_j)
\exp
\left\{  \sum_{i=1}^{N}
     \left(
          A  \bar{\phi}(\zeta_i )+ B  \zeta_i^2
     \right)
+F_N(\zeta_1,\cdots ,\zeta_N)
\right\}
\right] |0\rangle \! \rangle~,
\end{equation}
provided $F_N(\zeta_1,\cdots ,\zeta_N)$ satisfies
\begin{equation}
\partial_{\zeta_i}F_N (\zeta_{1}, \ldots, \zeta_{N})
 =\sum_{j\neq i}\frac{2}{\zeta_i-\zeta_j}~.
 \label{eq:diffeqforF}
\end{equation}
Thus we get
\begin{equation}
F_N(\zeta_1,\cdots ,\zeta_N)
=
2\sum_{i>j}\ln (\zeta_i-\zeta_j)~,
\end{equation}
and
\begin{equation}
|D_{N+} \rangle \! \rangle
=
\lambda_{N+} \int \prod_{i=1}^Nd\zeta_i 
\bigtriangleup_N^2 (\zeta_1,\cdots, \zeta_N )
\exp \left[ \sum_{i=1}^N
           \left( A \bar{\phi}(\zeta_i) +B \zeta_i^2 \right)
      \right]  |0\rangle \! \rangle~.
\end{equation}
Here $\bigtriangleup_N$ is the Vandermonde determinant. 

Notice that the integration measure 
\begin{equation}
\prod_{i=1}^Nd\zeta_i 
\bigtriangleup_N^2 (\zeta_1,\cdots, \zeta_N )
\end{equation}
coincides with that of the matrix models. 
This is natural if we regard $\zeta$ as the constant mode of tachyon 
on the D-brane. 
Since $\alpha$ can be considered as the length of the string, 
$\zeta$ may be identified with a constant tachyon mode on the 
boundary~\cite{Baba:2006rs}.
When there exist $N$ D-branes, 
the tachyon field becomes a matrix and we should consider $\zeta_i$ as 
its eigenvalues. 
Therefore we here encounter a matrix model of constant tachyons. 

$|D_{N+} \rangle \! \rangle$ can be considered as a state with $N$ D-branes. 
We can also construct a state with $N$ D-branes and $M$ ghost D-branes as
\begin{eqnarray}
|D_{N+,M-} \rangle \! \rangle
&\equiv&
\lambda_{N+,M-} \int 
\prod_{i=1}^Nd\zeta_i \prod_{\bar{\imath}=1}^M d\zeta_{\bar{\imath}}
\prod_{i>j}(\zeta_i-\zeta_j)^2 
\prod_{\bar{\imath}>\bar{\jmath}}
       (\zeta_{\bar{\imath}}-\zeta_{\bar{\jmath}})^2
\prod_{i,\bar{\jmath}}(\zeta_i-\zeta_{\bar{\jmath}})^{-2}
\nonumber
\\
& &
\quad
\times
\exp \left[ A
            \left(\sum_{i=1}^N\bar{\phi}(\zeta_i)
                 -\sum_{\bar{\imath}=1}^M\bar{\phi}(\zeta_{\bar{\imath}})
            \right)
            + B
            \left(\sum_{i=1}^N\zeta_i^2
                 -\sum_{\bar{\imath}=1}^M\zeta_{\bar{\imath}}^2
            \right)
      \right] |0\rangle \! \rangle~.
\label{eq:DN+M-}
\end{eqnarray}
This time the integration measure is that of the supermatrix model. 

Before closing this section, one comment is in order. 
It is possible to express the state $|D_{N+,M-} \rangle \! \rangle$ as
\begin{equation}
|D_{N+,M-} \rangle \! \rangle
\propto
\left( \int d\zeta\, \mathcal{V}_{D_{+}} (\zeta)
      \right)^{N}
\left( \int d\zeta' \, \mathcal{V}_{D_{-}} (\zeta')
      \right)^{M}
 | 0 \rangle \! \rangle
 ~.
\label{eq:vertex-like-op}
\end{equation}
Here $\mathcal{V}_{D_{\pm}}(\zeta)$ are of the form
\begin{equation}
\mathcal{V}_{D_{\pm}}(\zeta)
 = \bar{\mathcal{O}}_{D_{\pm}} (\zeta) \mathcal{O}_{D_{\pm}}(\zeta)~,
 \label{eq:ansatzVDp}
\end{equation}
where
$\bar{\mathcal{O}}_{D_{\pm}} (\zeta)$ are the operators given
in eq.(\ref{eq:solitonicoppm}) and $\mathcal{O}_{D_{\pm}}(\zeta)$
are defined as
\begin{equation}
\mathcal{O}_{D_{\pm}} (\zeta)
 = \exp \left[ \pm \int^{\infty}_{0} dr\,
                 \frac{e^{\zeta \alpha_{r}}}{\alpha_{r}}
                {}_{r}\langle v | \psi\rangle_{r} \right]~,
\end{equation}
with $|v\rangle$ satisfying
\begin{equation}
\int d'r \, {}_{r}\langle v | B_0 \rangle^{\epsilon}_{r}
 = - \frac{4}{A}~.
 \label{eq:condition-v}
\end{equation}
$\mathcal{V}_{D_{\pm}}(\zeta)$ 
look like vertex operators and 
may be considered as creation operators 
of D-branes and ghost D-branes. 
$|v\rangle$ can be any state as long as it satisfies 
eq.(\ref{eq:condition-v}). 
For example, 
$|v\rangle$ can be taken to be proportional to 
$| B_0 \rangle^{\epsilon}$ as in \cite{Baba:2006rs}.

\section{Disk Amplitudes}\label{sec:disk}          

Now that we have BRST invariant observables made from the boundary 
states, we would like to calculate the scattering amplitudes 
involving these operators and show that the amplitudes involving 
D-branes are reproduced. 
In particular, we would like to calculate the disk amplitudes 
in this paper.

\subsection{Three-point S-matrix elements}

Before going into the calculation of the disk amplitudes, 
it is instructive to recall how usual 
three-point S-matrix elements can be calculated
in the $OSp$ invariant string field theory~\cite{Baba:2007je}. 
Actually the calculation of the disk amplitudes goes in the same way 
as that for the three-point amplitudes. 
We also write down the space-time low energy effective action
of the $OSp$ invariant string field theory, which will be used 
to check the normalization and the sign of the amplitudes
involving D-branes. 

The S-matrix elements can be deduced from the correlation functions 
of the BRST invariant observables of the form~\cite{Baba:2007je}
\begin{equation}
\mathcal{O} (t,k)
= \int dr \, \frac{1}{\alpha_{r}}
  \, 
  {}_{r}\Bigl(
     {}_{C,\bar{C}} \langle 0 | \otimes
     {}_{X} \langle \mathrm{primary};k|
     \Bigr)
         |\Phi(t)\rangle_{r}~,
\label{eq:observable}
\end{equation}
where ${}_{C,\bar{C}} \langle 0 |$
and ${}_{X} \langle \mathrm{primary};k|$ denote
the BPZ conjugates of the Fock vacuum $|0\rangle_{C,\bar{C}}$
in the $(C,\bar{C})$ sector and a Virasoro primary state
$|\mathrm{primary};k \rangle_X$
in the $X^\mu$ sector $(\mu=1,\ldots,26)$ of the Hilbert space
for the worldsheet theory, respectively.
Here $k_{\mu}$ is the momentum eigenvalue of the state
$|\mathrm{primary};k\rangle_{X}$:
\begin{eqnarray}
 | \mathrm{primary};k  \rangle_{X}
 &=& |\overline{\mathrm{primary}} \rangle_{X}
  (2\pi)^{26} \delta^{26} (p -k )~,
\nonumber\\
{}_{X} \langle \mathrm{primary};k|
  &=& (2\pi)^{26}\delta^{26} (p+k)
   \, {}_{X} \langle \overline{\mathrm{primary}}|~.
\end{eqnarray}
$|\overline{\mathrm{primary}} \rangle_{X}$
denotes the non-zero mode part of $|\mathrm{primary};k\rangle_{X}$,
and we normalize it as
\begin{equation}
{}_{X} \langle \overline{\mathrm{primary}}|
         \overline{\mathrm{primary}} \rangle_{X} = 1~.
\end{equation}
The mass $M$ of the particle corresponding to the operator
$\mathcal{O}(t,k)$ can be read off from the relation
\begin{equation}
\left( L_{0} + \tilde{L}_{0} -2 \right)
  |\mathrm{primary};k\rangle_{X} \otimes |0\rangle_{C,\bar{C}}
= \left( k^{2} + 2i \pi_{0} \bar{\pi}_{0} + M^{2}  \right)
   |\mathrm{primary};k\rangle_{X} \otimes |0\rangle_{C,\bar{C}}~.
\end{equation}
Since we consider correlation functions, the primary states
introduced here are off-shell in general,
i.e.\ $k^{2} + M^{2} \neq 0$.
For later use,
we introduce the on-shell primary states
$
|\mathrm{primary};\mathbf{k} \rangle_{X}
 = \left. |\mathrm{primary};k \rangle_{X}
   \right|_{k^{2} + M^{2} =0}$~,
where $\mathbf{k}$ denotes the spatial $25$-momentum.

By using the canonical commutation relation (\ref{eq:ccr}),
the lowest order contribution of
the three-point correlation function for the observables
$\mathcal{O}_{r}(t_{r},k_r)$ $(r=1,2,3)$ $(t_1 > t_2 > t_3)$
with mass $M_{r}$ is evaluated as
\begin{eqnarray}
\lefteqn{
 \bigg\langle \!\! \bigg\langle
   \mathcal{O}_{1}(t_{1},k_1) \mathcal{O}_{2}(t_{2},k_2)
   \mathcal{O}_{3}(t_{3},k_3)
 \bigg \rangle \!\! \bigg\rangle
}
\nonumber\\
&&
= 
  \left[ \int^{t_{2}}_{t_{3}} dT
         \prod_{s=1}^{2} \left(
           -\int^{0}_{-\infty} \frac{d\alpha_{s}}{2} \right)
           \int^{\infty}_{0} \frac{d\alpha_{3}}{2}
         + \int^{t_{1}}_{t_{2}} dT
            \left( - \int^{0}_{-\infty} \frac{d\alpha_{1}}{2} \right)
             \prod_{s=2}^{3}
               \left( \int^{\infty}_{0} \frac{d\alpha_{s}}{2}
               \right)
  \right]
\nonumber\\
&& \ \quad
    \times   4ig 
    \prod_{r'=1}^{3} \left(
     \int \frac{d^{26}p_{r'}}{(2\pi)^{26}} id\bar{\pi}_{0}^{(r')}
      d \pi_{0}^{(r')}  \right)
 \nonumber\\
&& \hspace{3.5em}
    \times
     \left\langle V^{0}_{3}(1,2,3) \right|
    \prod_{r=1}^{3} \left[
     e^{-i \frac{|T-t_r|}{|\alpha_r|} \left(p_{r}^{2} + M_{r}^{2}
               +2i\pi^{(r)}_{0}\bar{\pi}^{(r)}_{0} \right)}
       \left(
       |\mathrm{primary}_{r}; k_{r} \rangle_X
        \otimes |0\rangle_{C,\bar{C}}
       \right)_{r} 
       \right].~~~
\label{eq:mathcalO123}
\end{eqnarray}
We can readily integrate over $\alpha_3, \pi_0^{(3)}, \bar{\pi}_0^{(3)}, p_{1}$ and 
$p_{2}$, using the delta functions. 
In order to obtain the S-matrix elements, 
we need to look for the on-shell poles for the external momenta.
The singular behavior at $k_2^2 + M_2^2 =0$ comes from 
the region $\alpha_2\sim 0$ in the integration over 
$\alpha_2$~\cite{Baba:2007je}. 
Therefore we should consider the limit $\alpha_2\rightarrow 0$ in the 
three-string vertex
$\left\langle V^{0}_{3}(1,2,3) \right|$.
In this limit, 
the complicated expression (\ref{eq:mathcalO123})
involving three Hilbert spaces for strings $1$, $2$ and $3$ can 
be simply described in terms of the vertex operator as follows:
\begin{eqnarray}
&&
 \bigg\langle \!\! \bigg\langle
   \mathcal{O}_{1}(t_{1},k_1) \mathcal{O}_{2}(t_{2},k_2)
   \mathcal{O}_{3}(t_{3},k_3)
 \bigg \rangle \!\! \bigg\rangle
\nonumber\\
&&
\sim \frac{1}{k_2^2 +M_2^2} \, 4ig
    \int_{t_3}^{t_1} dT\,
            \int^{0}_{-\infty} \frac{d\alpha_{1} }{2 \alpha_{1}}
            \int i d\bar{\pi}^{(1)}_{0} d\pi^{(1)}_{0} \,
  \frac{1}{\alpha_{1}} \,
  e^{-i \frac{t_1-T}{-\alpha_1} \left(k_{1}^{2} + M_{1}^{2}
               +2i\pi^{(1)}_{0}\bar{\pi}^{(1)}_{0} \right) }
 \nonumber\\
&& \qquad \quad
   \times
       e^{-i \frac{T- t_3 }{-\alpha_1} \left(k_{3}^{2} + M_{3}^{2}
               +2i\pi^{(1)}_{0}\bar{\pi}^{(1)}_{0} \right) }
     \int \frac{d^{26}p}{(2\pi)^{26}}
     \,
      {}_{X} \langle \mathrm{primary}_{1};k_{1}|
      \mathcal{V}_{2} (\mathbf{k}_{2}) | \mathrm{primary}_{3};k_{3}
      \rangle_{X}~,~~~~
\label{eq:green3}
\end{eqnarray}
where $\mathcal{V}_{2}(\mathbf{k}_{2})$ denotes the vertex operator
corresponding to the primary state
$|\mathrm{primary}_{2};\mathbf{k}_{2}\rangle_{X}$
on the mass-shell
associated with the observable $\mathcal{O}_2$.
After the integration over $\pi_0^{(1)}$ and $\bar{\pi}_0^{(1)}$,
eq.(\ref{eq:green3}) becomes
\begin{eqnarray}
&=& \frac{1}{k_2^2 +M_2^2} (-4g)
  \int^{\infty}_{0} dT'
  \int_{0}^{\infty} dT''
     e^{-i T' \left(k_{1}^{2} + M_{1}^{2} \right) }
     e^{-i T'' \left(k_{3}^{2} + M_{3}^{2} \right) } 
\nonumber\\
&&
\qquad \qquad \qquad \times 
\int \frac{d^{26}p}{(2\pi)^{26}}
     \,
      {}_{X} \langle \mathrm{primary}_{1};k_{1}|
      \mathcal{V}_{2} (\mathbf{k}_{2}) | \mathrm{primary}_{3};k_{3}
      \rangle_{X}  
\nonumber\\
&\sim&  \prod_{r=1}^{3} \left(\frac{1}{k_r^2 + M_r^2} \right)
 4g \int \frac{d^{26}p}{(2\pi)^{26}}
     \,
      {}_{X} \langle \mathrm{primary}_{1};\mathbf{k}_{1}|
      \mathcal{V}_{2} (\mathbf{k}_{2}) 
      | \mathrm{primary}_{3};\mathbf{k}_{3}
      \rangle_{X}~.
\label{eq:3-pointGF}
\end{eqnarray}
Here we have changed the integration variables from
$T$ and $\alpha_{1}$ to $T'$ and $T''$,
where $T'= \frac{t_1 - T} {-\alpha_1}$ 
and $T''= \frac{T - t_3 } {-\alpha_1}$.
Carrying out the Wick rotation to make the space-time signature
Lorentzian,
we can see that the lowest order contribution
to the S-matrix element for this process is
\begin{equation}
S=  4ig  
 \int \frac{d^{26}p}{(2\pi)^{26}}
     \,
      {}_{X} \langle \mathrm{primary}_{1};\mathbf{k}_{1}|
      \mathcal{V}_{2} (\mathbf{k}_{2}) 
      | \mathrm{primary}_{3};\mathbf{k}_{3}
      \rangle_{X}~.
\label{eq:3-pointSmatrix}
\end{equation}

In following subsections,
we will discuss the normalization and the sign of the disk amplitudes. 
In doing so, we need 
the space-time low energy effective action
for the tachyon $T(x)$ and the graviton $h_{\mu\nu}(x)$.
Let us calculate the S-matrix elements 
for processes involving only tachyons and gravitons.
The primary states corresponding to these particles are 
\begin{eqnarray}
|\mathrm{primary}_{r}; k_{r} \rangle_X &=&
\left\{
    \begin{array}{ll}
       |0\rangle_X \, (2\pi)^{26} \delta^{26} (p-k_{r})
           & \mbox{for the tachyon} 
     \\
        e_{r,\mu\nu}(k_{r}) \alpha_{-1}^{\mu} 
                \tilde{\alpha}_{-1}^{\nu} |0\rangle_X
                \, (2\pi)^{26}\delta^{26} (p-k_{r})
           & \mbox{for the graviton}
    \end{array}
\right.,
\label{eq:primary}
\end{eqnarray}
where $|0\rangle_X$ denotes the Fock vacuum for the $X^\mu$ sector
and $e_{r,\mu\nu}(k_{r})$ denotes 
the polarization of the asymptotic graviton state
with momentum $k_{r,\mu}$.
The polarization $e_{r,\mu\nu}(k_{r})$ satisfies the following relations:
\begin{equation}
e_{r,\mu\nu}=e_{r,\nu\mu}~,\quad
\eta^{\mu\nu}e_{r,\mu\nu}=0~,\quad
 k_{r}^{\mu} e_{r,\mu\nu}=0~,\quad
e_{r,\mu\nu} \, e_{r}^{\mu\nu} =1~.
\label{eq:polarization}
\end{equation}
The vertex operators appearing in eq.(\ref{eq:green3}) are
\begin{eqnarray}
\mathcal{V}_{r} (\mathbf{k}_{r})
 &=& \maru \, e^{i k_{r,\mu} X^{\mu}}(0) \, \maru
\label{eq:vertexT}
\end{eqnarray}
for the tachyon and
\begin{eqnarray}
\mathcal{V}_{r} (\mathbf{k}_{r})
 &=& - e_{r,\mu\nu}(k_{r}) \, 
    \maru\, \partial X^\mu \bar{\partial} X^\nu 
                e^{i k_{r,\lambda} X^{\lambda}}(0) \, \maru
\nonumber\\
&=& e_{r,\mu\nu} (k_{r}) \,
    \maru  \left( p^\mu + \sum_{n \neq 0} \alpha_n^\mu \right) 
                  \left( p^\nu 
                     + \sum_{m \neq 0} \tilde{\alpha}_m^\nu \right) 
               e^{i k_{r,\lambda} X^{\lambda} (0)}
   \,\maru
\label{eq:vertexh}
\end{eqnarray}
for the graviton. In these equations,
$\, \maru~\maru\, $ denotes 
the normal ordering of the oscillators
and $0$ in the arguments of the operators indicates the
origin $(\tau,\sigma)=(0,0)$ of the worldsheet.

Plugging eqs.(\ref{eq:primary}),
(\ref{eq:vertexT}) and (\ref{eq:vertexh})
into eq.(\ref{eq:3-pointSmatrix}), we obtain 
three-point S-matrix elements for tachyons and gravitons:
\begin{eqnarray}
S_{TTT} 
&=& 4 ig
       \, (2\pi)^{26}\delta^{26}(k_1+k_2+k_3)~,
\nonumber\\
S_{TTh}
&=& ig \, e_{3,\mu\nu} k_{12}^\mu k_{12}^\nu
\, (2\pi)^{26}\delta^{26}\left(k_1 + k_2 + k_3\right)~,
\nonumber\\
S_{hhh} 
&=& ig \, 
e_{1,\mu\nu} e_{2,\alpha\beta}
e_{3,\gamma\delta}
      \, T^{\mu\alpha\gamma} T^{\nu\beta\delta}
\, (2\pi)^{26}\delta^{26}(k_1 + k_2 + k_3 )~,
\label{eq:s_ggg}
\end{eqnarray}
where the subscripts $T$ and $h$ denote the tachyon and the graviton
respectively and
\begin{eqnarray}
k^{\mu}_{rs}
 &=& k^{\mu}_r-k^{\mu}_s~,
\nonumber\\
T^{\mu\alpha\gamma}
&=& \eta^{\mu\alpha}k_{12}^\gamma
  + \eta^{\alpha\gamma}k_{23}^\mu  
  + \eta^{\gamma\mu}k_{31}^\alpha
  +\frac{1}{4}k_{23}^\mu 
                        k_{31}^\alpha k_{12}^\gamma~.
\end{eqnarray}
Eq.(\ref{eq:s_ggg}) coincide with the results
in the light-cone gauge string field theory.

We can reproduce
the results obtained in 
eq.(\ref{eq:s_ggg}) from the following space-time effective action
for the metric $G_{\mu\nu}(x)$ and the tachyon field $T(x)$,
\begin{eqnarray}
S=\frac{1}{2\kappa^2} \int d^{26}x \sqrt{-G} R
+ \int d^{26}x\sqrt{-G}
\left(-\frac{1}{2}G^{\mu\nu}\partial_\mu T \partial_\nu T
      +T^2 +\frac{2g}{3} T^3\right)
\nonumber\\ 
+\mbox{higher derivative terms}~,
\label{eq:s_eff}
\end{eqnarray}
by expanding the metric $G_{\mu\nu}(x)$ around
the flat metric $\eta_{\mu\nu}$ as
\begin{equation}
G_{\mu\nu}(x) = \eta_{\mu\nu} +2\kappa 
   h_{\mu\nu}(x)~.
\label{eq:metric_expand}
\end{equation}
We find that the gravitational coupling constant $\kappa$ 
is related to the string coupling $g$ as
\begin{equation}
\kappa=2g~.
\label{eq:kappa-g}
\end{equation}
%

\subsection{Disk amplitudes}

Now let us turn to the disk amplitudes. 
We evaluate the disk amplitude with two external closed string tachyons 
in the presence of one soliton, as an example.
We show that our results coincide with 
those for a (ghost) D-brane in string theory.
Using these disk amplitudes, we determine which of the states
$|D_{\pm} \rangle\!\rangle$ corresponds to the D-brane.

Since $|D_{\pm} \rangle\!\rangle$ is a BRST invariant state, 
we may be able to calculate the amplitudes involving D-branes 
by starting from the correlation function 
\begin{equation}
\langle \! \langle 0|\mathrm{T}
    \mathcal{O}_1 (t_1) \cdots \mathcal{O}_N (t_N) 
    |D_{\pm} \rangle \! \rangle~.
\label{eq:Dgreensfunction}
\end{equation}
Indeed, from $|D_{\pm} \rangle \! \rangle$ we get insertions 
of the boundary states and the worldsheets with boundaries are generated. 
However, because the formulation of the theory is similar to the light-cone 
field theory, we cannot generate the worldsheets without any external 
line insertions by considering
\begin{equation}
\langle \! \langle 0|D_{\pm} \rangle \! \rangle~.
\label{eq:vacuum}
\end{equation}
Such vacuum amplitudes are constants. 
Especially the cylinder amplitudes are constants which do not depend 
even on the coupling constant $g$. 
Therefore they can be considered to be included in the 
definition of the unknown constant $\lambda_{\pm}$.
If we replace the bra $\langle \! \langle 0|$
by $\langle \! \langle D_{\pm}|$ in eq.(\ref{eq:vacuum}),
we get worldsheets without any external line insertions. 
This is calculated in \cite{Baba:2006rs}. 
But the result is a constant and
cannot be distinguished from $\lambda_{\pm}$.

In order to normalize the correlation function (\ref{eq:Dgreensfunction}), 
we divide it by the vacuum amplitude 
as in the usual field theory, 
and consider
\begin{eqnarray}
&&\bigg\langle \!\! \bigg\langle 
\mathcal{O}_1 (t_1) \cdots \mathcal{O}_N (t_N) 
\bigg\rangle \!\! \bigg\rangle_{D_{\pm}}
=
\frac{\langle \! \langle 0|\mathrm{T}
    \mathcal{O}_1 (t_1) \cdots \mathcal{O}_N (t_N) 
    |D_{\pm} \rangle \! \rangle}
{\langle \! \langle 0|D_{\pm} \rangle \! \rangle }
~.
\end{eqnarray}
Therefore, starting from this normalized correlation function, 
we can calculate the amplitudes in the usual way. 

Now let us calculate correlation functions 
for two closed string tachyons in the presence of the soliton,
to obtain the S-matrix elements.
The correlation function to be calculated is
\begin{eqnarray}
&&\bigg\langle \!\! \bigg\langle 
\mathcal{O}_1^T (t_1,k_{1}) \mathcal{O}_2^T (t_2,k_{2}) 
\bigg\rangle \!\! \bigg\rangle_{D_{\pm}}
=
\frac{\langle \! \langle 0|
    \mathcal{O}_1^T (t_1,k_{1}) \mathcal{O}_2^T (t_2,k_{2}) 
    |D_{\pm} \rangle \! \rangle}
{\langle \! \langle 0|D_{\pm} \rangle \! \rangle }
~.
\label{eq:2pointDbrane}
\end{eqnarray}
Here $\mathcal{O}_r^T$ is the observable corresponding to the tachyon state, 
and $t_1 > t_2$.
The lowest order contributions to this correlation
function give the propagator and tadpole for the tachyon.
The $\mathcal{O}(g)$ term is what we should look at. 

In perturbation theory, 
$| D_{\pm} \rangle \! \rangle$ can be recast into a more tractable 
form as follows. 
In the integrand (\ref{eq:solitonicoppm}) of the integration
(\ref{eq:solitonicstatepm}), 
the factor
\begin{equation}
\exp\left[\pm \frac{(2\pi)^{13}\epsilon^{2}
                        \left(-\ln \epsilon\right)^{\frac{p+1}{2}}}
                   {16 \left(\frac{\pi}{2} \right)^{\frac{p+1}{2}} 
                         \sqrt{\pi} g }  \zeta^{2}
         \right]
  \label{eq:gaussian}
\end{equation}
becomes the most dominant perturbatively.
Therefore, we carry out the saddle point approximation to obtain
\begin{eqnarray}
|D_{\pm}  \rangle\!\rangle
\simeq \lambda'_{\pm}  \exp\left[ \pm   
          \frac{(2\pi)^{13}}
               {(8\pi^2)^{ \frac{p+1}{2} } \sqrt{\pi} } 
       \int_{-\infty}^{0} \frac{dr}{\alpha_r}
        {}_r^{\epsilon}\langle B_0 
        | \bar{\psi} \rangle_r \right]
  |0\rangle\!\rangle~,
\label{eq:saddle}
\end{eqnarray}
where $\lambda'_{\pm}$ is given as
\begin{equation}
\lambda'_{\pm}
\equiv
 \sqrt{\mp\frac{16 \left( \frac{\pi}{2}\right)^{\frac{p+1}{2} } 
                                                \pi^{\frac{3}{2}} g }
          { (2\pi)^{13} \epsilon^{2} (-\ln \epsilon)^{\frac{p+1}{2}} } }
   \; \lambda_{\pm}~.
\label{eq:lambdaprime}
\end{equation}
Notice that for $|D_{+}  \rangle\!\rangle$ the exponent
of the Gaussian factor (\ref{eq:gaussian}) has the wrong sign, 
which makes the factor in front of $\lambda_{+}$
in eq.(\ref{eq:lambdaprime}) pure imaginary.
This is a sign of instability.

\begin{figure}[htbp]
\vspace{1.5\baselineskip}
\begin{center}
\begin{picture}(320,140)(0,30)
\put(0,30){
\includegraphics[width=11em,clip]{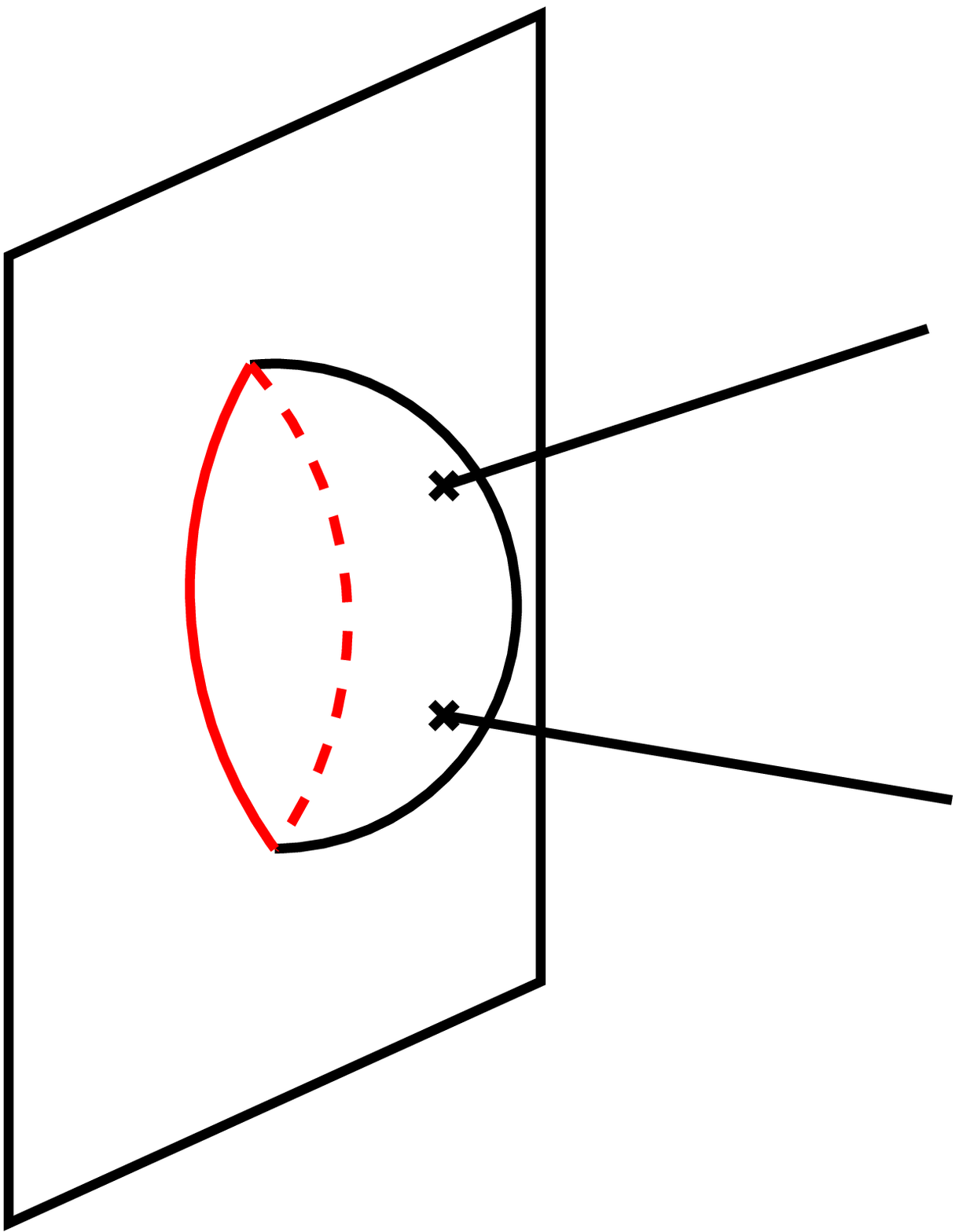}
}
\put(200,30){
\includegraphics[width=11em,clip]{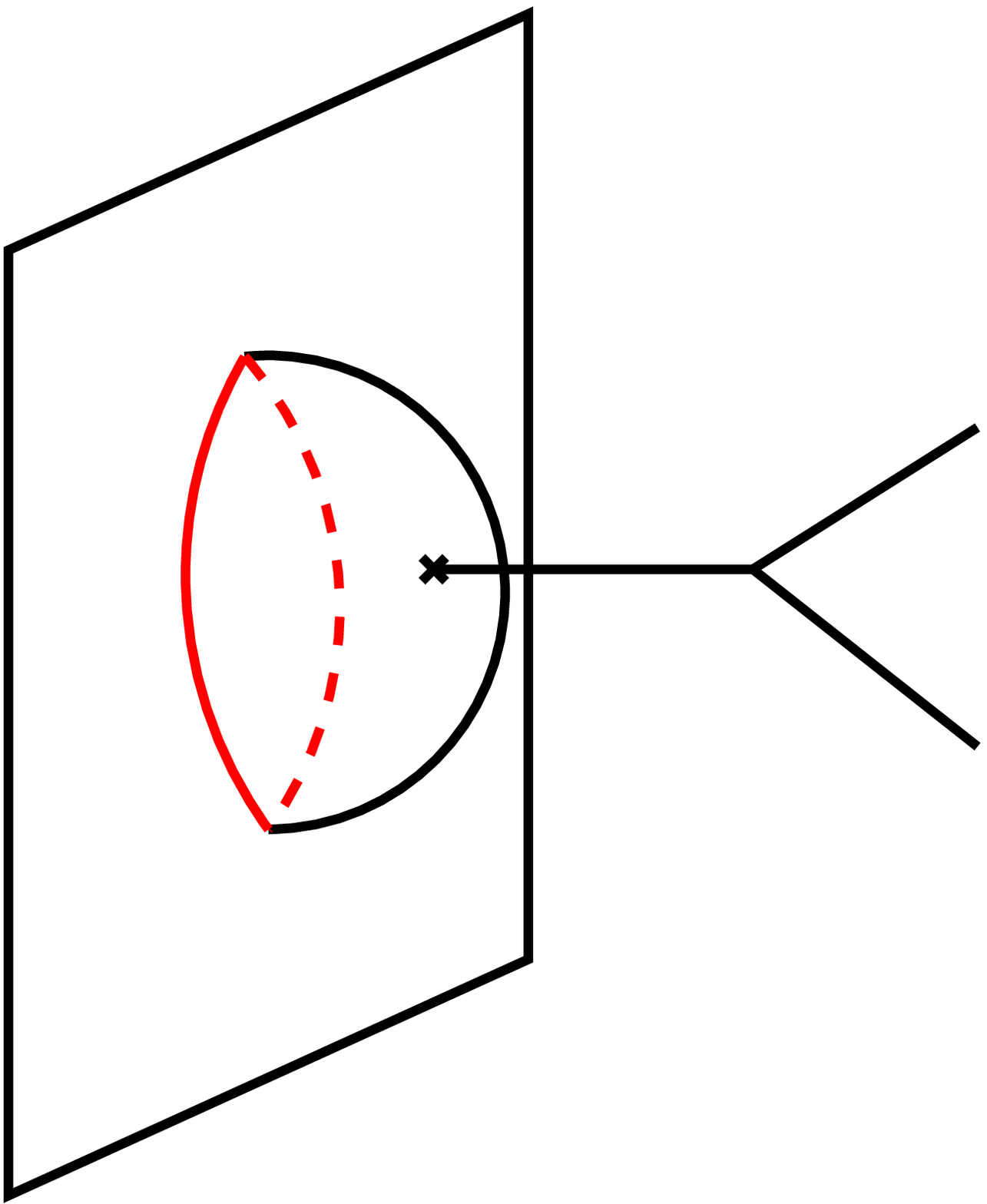}
}
\put(70,15){(a)}
\put(270,15){(b)}
\end{picture}
\end{center}
\begin{quote}
\caption{(a) The worldsheet diagram of 
the two-tachyon disk amplitudes.
   (b) The worldsheet diagram that contributes to the pole of intermediate
 closed string states.}
\label{fig:disk}
\end{quote}
\end{figure}

Then 
the $\mathcal{O}(g)$ term can be given as
\begin{eqnarray}
&&G_{TTD_{\pm}} (k_1,k_2)
\nonumber\\
&& =  
\left[
\int_{t_3}^{t_2} dT
\prod_{s=1}^{2}
  \left(
     - \int _{-\infty}^{0} \frac{d\alpha_s}{2} 
  \right)
   \left(
      \int ^{\infty}_{0} \frac{d\alpha_3}{2} 
  \right)  
+ \int_{t_2}^{t_1} dT
   \left(
       - \int _{-\infty}^{0} \frac{d\alpha_1}{2} 
  \right)
\prod_{s=2}^{3}
  \left(
      \int ^{\infty}_{0} \frac{d\alpha_s}{2} 
  \right)
\right]
\nonumber\\
&&  \quad \times
 \frac{\pm 4ig(2\pi)^{13}}
             {(8\pi^2)^{\frac{p+1}{2}} \sqrt{\pi} }
\prod_{r'=1}^{3} \left(
  \int \frac{d^{26}p_{r'}}{(2\pi)^{26}} id\bar{\pi}^{(r')}_{0}
       d\pi_{0}^{(r')} \right)
\left\langle V_3^0(1,2,3) \right|
  \nonumber\\
&& \quad  \times
\prod_{r=1}^2
\left(
    e^{-i\frac{|T- t_r|} {|\alpha_r|}
        \left(p_r^2 + 2i\pi_0^{(r)} \bar{\pi}_0^{(r)} -2 \right) } 
         | 0\rangle_r
(2\pi)^{26} \delta^{26} (p_r-k_r)
\right)
    e^{-i \frac{T - t_{3}}{\alpha_3}
        \left(L_0^{(3)}+\tilde{L}^{(3)}_0-2\right)  }
         |B_0\rangle_3^{\epsilon}~,
\label{eq:s_ttd}
\end{eqnarray}
where $t_{3}$ $(<t_1,t_2)$ is the proper time of the solitonic state.
In what follows, we will show that
this correctly provides the contribution of 
the disk attached to the (ghost) D-brane
corresponding to our solitonic states $|D_{\pm} \rangle\!\rangle$.
The worldsheet diagram of this process
is depicted in Fig.~\ref{fig:disk}(a).

Eq.(\ref{eq:s_ttd}) is quite similar to eq.(\ref{eq:mathcalO123}) and 
can be calculated in the same way. 
Looking for the singular behavior at $k_2^2 -2 =0$, 
we can get
\begin{eqnarray}
\lefteqn{
G_{TTD_{\pm}}(k_1, k_2)
}
\nonumber\\
&\sim& \frac{1}{(k_2^2 - 2)}
        \frac{\pm 4ig(2\pi)^{13}}
             {(8\pi^2)^{\frac{p+1}{2}} \sqrt{\pi} } 
  \int_{t_3}^{t_1} dT \,
     \int_{-\infty}^{0} \frac{d\alpha_1}{2 \alpha_1}
                        \int i d\bar{\pi}_0^{(1)} d\pi_0^{(1)}
  \, \frac{1}{\alpha_1}\,
  e^{-i \frac{T - t_1}{-\alpha_1} 
             \left(k_1^2 + 2i \pi_0^{(1)} \bar{\pi}_0^{(1)} -2 \right) }
 \nonumber\\
&& \quad \times
  \int \frac{d^{26}p}{(2\pi)^{26}}
  (2\pi)^{26} \delta^{26} (p+k_{1})
  \,  {}_X \langle 0| \, \maru \, e^{ik_{2,\mu} X^{\mu}}(0) \, \maru\; 
        e^{-i \frac{T -t_3}{-\alpha_1} 
              \left( L_0^{X} + \tilde{L}_0^{X}
                     + 2i \pi_0^{(1)} \bar{\pi}_0^{(1)} -2\right)}
  |B_0 \rangle_X
 \nonumber\\
&=& \frac{1}{(k_2^2 - 2)}
        \frac{\pm 4ig(2\pi)^{13}}
             {(8\pi^2)^{\frac{p+1}{2}} \sqrt{\pi} } 
      \; i \int_{0}^{\infty} dT' 
   \int^{\infty}_{0} dT''
  e^{-i T' \left(k_1^2 -2 \right) }
\nonumber\\
&&\quad \times
  \int \frac{d^{26}p}{(2\pi)^{26}}
   (2\pi)^{26} \delta^{26} (p+k_{1})
  {}_X \langle 0|\, \maru \, e^{ik_{2,\mu} X^{\mu} }(0) \,\maru \;
        e^{-i T''
              \left( L_0^{X} +\tilde{L}_0^{X} -2 \right) }
  |B_0 \rangle_X
\nonumber\\
&\sim& 
  \frac{1}{k_1^2-2} \frac{1}{k_2^2 - 2}
  \, \frac{\pm4ig(2\pi)^{13}}
             {(8\pi^2)^{\frac{p+1}{2}} \sqrt{\pi}  } 
\nonumber\\
&& \quad \times
\int \frac{d^{26}p}{(2\pi)^{26}}
  (2\pi)^{26} \delta^{26} (p+k_{1})
{}_X\langle 0| \,\maru\, e^{ik_{2,\mu} X^{\mu}}(0) \,\maru\;
\frac{-i}{L_0^X +\tilde{L}_0^X -2 }
|B_0 \rangle_X ~,
\end{eqnarray}
where $ L_0^{X}$ and $\tilde{L}_0^{X}$
are the zero-modes of the Virasoro generators
 and $|B_0\rangle_X$ is
the boundary state in the $X^\mu$ sector, respectively:
\begin{eqnarray}
L_0^X &=& 
 \frac{1}{2} p^2
  + \sum_{\mu \in \mathrm{N}, \mathrm{D}} \sum_{n=1}^\infty 
       \alpha_{-n}^{\mu} \alpha_{n \mu}~,
\quad
\tilde{L}_{0}^X =
 \frac{1}{2} p^{2}
   + \sum_{\mu \in \mathrm{N},\mathrm{D}} \sum_{n=1}^{\infty}
       \tilde{\alpha}_{-n}^{\mu} \tilde{\alpha}_{n\mu}~,
\nonumber\\
|B_0 \rangle_{X}
&=& \exp\left[ - \sum_{\mu,\nu \in \mathrm{N},\mathrm{D}}\,
                 \sum_{n=1}^{\infty} \frac{1}{n} \alpha_{-n}^\mu 
                 \tilde{\alpha}_{-n}^\nu D_{\mu\nu} 
        \right] |0\rangle_X
    (2\pi)^{p+1} \delta_{\mathrm{N}} ^{p+1}(p)~.
\end{eqnarray}
Carrying out the Wick rotation, 
we find that the S-matrix element for this process is
\begin{equation}
S_{TTD_{\pm}}
=  \frac{\pm 4ig (2\pi)^{13} }
           { (8\pi^2)^{\frac{p+1}{2} } \sqrt{\pi} }
\int \frac{d^{26}p}{(2\pi)^{26}}
  (2\pi)^{26} \delta^{26} (p+k_{1})
{}_X\langle 0| \,\maru\, e^{ik_{2,\mu} X^{\mu}}(0) \,\maru\;
\frac{1}{L_0^X +\tilde{L}_0^X -2 }
|B_0 \rangle_X ~,
\label{eq:s_ttd2}
\end{equation}
where the momenta $k_{r,\mu}$ $(r=1,2)$ are subject to the on-shell
condition for the tachyon: $k_{r}^{2}=2$.
It is clear that the amplitude is proportional to the 
usual disk amplitude.

It is straightforward to generalize the above calculations for 
other closed string states, just by replacing the state and 
the vertex operator. 
Also it is quite obvious that we can reproduce the disk amplitudes 
with more than two external lines. 
In order to consider the situation in which there are more than 
one solitons, 
we should replace $|D_{\pm}\rangle\!\rangle$ by 
$|D_{N+,M-} \rangle \! \rangle$. 
The leading order contribution in perturbation theory is
from $\zeta_i=\zeta_{\bar{\imath}}=0$ in eq.(\ref{eq:DN+M-}) 
and we obtain the S-matrix element as 
$S_{TTD_+}$ in eq.(\ref{eq:s_ttd2}) multiplied by $N-M$. 

We can also replace the bra $\langle \! \langle 0|$ in 
eq.(\ref{eq:2pointDbrane}) by the solitonic states. 
By doing so, we introduce more solitons and 
it is easy to see that 
the disk amplitudes 
are multiplied by the total number of D-branes minus that 
of ghost D-branes. 
Therefore, it is now clear that
we considered situations with even number 
of solitons in \cite{Baba:2006rs}, 
by taking the bra and the ket to be hermitian conjugate 
to each other. 
In this paper, considering that the vacuum amplitudes are 
included in the definitions of $\lambda_\pm$, 
we can realize more general situations. 

\subsection{D-brane and ghost D-brane states}
\label{sec:D-ghostD}

Let us check if the disk amplitude~(\ref{eq:s_ttd2}) 
has the correct normalization. 
At the on-shell pole of an intermediate closed string state 
$|\mathrm{primary};k\rangle_{X}$, 
it is factorized as 
\begin{eqnarray}
&&
S_{TTD_{\pm}} \sim \int \frac{d^{26}k}{(2\pi)^{26}}
\left[4ig 
\int \frac{d^{26}p'}{(2\pi)^{26}}
 (2\pi)^{26}  \delta^{26}(p'+k_{1})
 {}_X \langle 0| \, \maru\, e^{ik_{2,\mu} X^{\mu}}(0) \,\maru\, 
                      |\mathrm{primary};k\rangle_{X}
\right]
\nonumber\\
&&\qquad \qquad \qquad \qquad
\times
\frac{-i}{k^2 + M^2}
\times
\left[
\frac{\pm i(2\pi)^{13} }
           {(8\pi^2)^{\frac{p+1}{2} } \sqrt{\pi}}
\int  \frac{d^{26}p}{(2\pi)^{26}}
{}_X \langle \mathrm{primary};-k| B_0\rangle_X
\right]~,
\label{eq:disk-pole}
\end{eqnarray}
where $M$ denotes the mass of the state. 
Since the D-brane can be considered as 
a source of closed string states,
the low energy effective action
should have source terms at $x^i =0 \; (i \in \mathrm{D})$
due to the presence of solitons.
From eq.(\ref{eq:disk-pole}), we can read off the source terms 
as 
\begin{equation}
S'_{\pm}= \pm \frac{(2\pi)^{13} }{(8 \pi^2)^\frac{p+1}{2} \sqrt{\pi} }
\int d^{26} x\,
      \prod_{i\in \mathrm{D}} \delta(x^i)
      \, \left[\, T(x)
             - 2 \sum_{\mu,\nu \in\mathrm{N}} h_{\mu\nu}(x) \eta^{\mu\nu}
       + \cdots \, \right] ~,
\label{eq:source}
\end{equation}
where the ellipsis denotes
the contribution from the states other than the tachyon $T(x)$ and
the graviton $h_{\mu\nu}(x)$.
This can be compared with the 
DBI action for a flat D$p$-brane located at
$x^{i}=0$ $(i\in \mathrm{D})$:
\begin{equation}
S_{p} = - \tau_{p} \int d^{26}x
          \prod_{i\in \mathrm{D}} \delta (x^{i})
          \sqrt{ \, - \det_{\mu,\nu \in \mathrm{N}} G_{\mu\nu} (x)
                \rule{0em}{0.85em}}~,
\label{eq:DBIaction}
\end{equation}
where $\tau_{p}$ is the D$p$-brane tension
in bosonic string theory defined
as~\cite{Dai:1989ua}\cite{Polchinski:1996fm}\footnote{
In this paper, we use the units in which $\alpha'=2$.}
\begin{equation}
\tau_{p} = \frac{\sqrt{\pi}}{16 \kappa} (8\pi^{2})^{\frac{11-p}{2}}~.
\end{equation}
Using eq.(\ref{eq:metric_expand}) we can expand $S_{p}$ in terms of 
$h_{\mu\nu}(x)$,
and obtain the source term for $h_{\mu\nu}(x)$
which coincides with that in $S_+^\prime$ in eq.(\ref{eq:source}).
Therefore 
the disk amplitude $S_{TTD_{+}}$ coincides with that for a D-brane 
and $S_{TTD_{-}}$ coincides with that for a ghost D-brane. 

Hence we should identify $|D_{+}\rangle\!\rangle$ with the state with 
one D-brane and 
$|D_{-}\rangle\!\rangle$ with the state with one ghost D-brane. 
This identification is quite consistent. 
D-branes in bosonic string theory are unstable due to the lack 
of the RR-charge and 
the soliton corresponding to the state $|D_{+}\rangle \! \rangle$
is also unstable, as was mentioned below eq.(\ref{eq:lambdaprime}).

\section{Discussion}\label{sec:discussion}

In this paper, we construct solitonic states corresponding to D-branes 
and ghost D-branes and check that the disk amplitudes coincide with 
the usual string theory results. 
These solitonic states are BRST invariant in the leading order of 
$\epsilon$. 
Since the BRST variation in eq.(\ref{eq:totalder}) is 
of order $\epsilon^{-2}(-\ln \epsilon )^{-\frac{p+1}{2}}$, 
higher order corrections do not go to $0$ in the limit $\epsilon \to 0$. 
For $p\neq -1$, the correction terms are of order 
$\epsilon^{-2}(-\ln \epsilon )^{-\frac{p+1}{2}-n}~(n>0)$ and 
for $p=-1$, the next leading term is of order $\epsilon^0$. 
It might be possible to prove that by modifying the exponent of 
$\bar{\mathcal{O}}_{D\pm}(\zeta)$ as 
\begin{equation}
 \exp \left[ \pm A\bar{\phi}(\zeta )
         \pm B\zeta^2 +(\mathrm{terms~higher~order~in~} \epsilon )
        \right]~,
\end{equation}
it becomes BRST invariant. 
As is clear from the calculation of the disk amplitudes, 
the higher order terms do not contribute to the amplitudes 
in the limit $\epsilon\to 0$. 
Of course, we need to examine the form of the BRST transformation 
to show that this actually happens. 
We do not try doing so, because here we are dealing with bosonic strings 
and we are destined to have insurmountable divergences any way. 
Hopefully, we may be able to show the BRST invariance more completely 
in the superstring case. 

The calculation of the disk amplitudes goes in the same way as that 
in the usual amplitudes. 
Open string external lines may be introduced by deforming the boundary 
state by the marginal operators corresponding to the open string 
vertex operators. 
It is an intriguing problem to  examine if the higher order open 
string amplitudes are reproduced correctly. 
Another problem is to calculate the open string amplitudes without 
closed string insertions. 

The variables $\zeta$ in the definition of the solitonic states 
can be regarded as constant tachyon. 
They are conjugate to the $\alpha$ in the $OSp$ invariant string field 
theory. 
Therefore somehow a part of the open string modes is incorporated in 
the formulation of the closed string field theory. 
It may be possible to generalize this to other modes of open strings.

\section*{Acknowledgements}

We would like to thank I.~Kishimoto, T.~Saitou and Y.~Satoh
for discussions and comments.
This work was supported in part by Grant-in-Aid for
Young Scientists (B) (19740164) from
the Ministry of Education, Culture, Sports, Science and
Technology (MEXT), and Grant-in-Aid for
JSPS Fellows (19$\cdot$1665).

\newpage
\appendix

\section{$OSp$ Invariant String Field Theory}\label{sec:conventions}

In this appendix, we summarize the formulation of 
the $OSp$ invariant string field theory.

\subsubsection*{variables}
The coordinate variables on the worldsheet in the $OSp$ invariant 
string field theory are the 
$OSp(26|2)$ vector
$X^{M}=\left( X^\mu ,C,\bar{C}\right)$,
where $X^{\mu}~(\mu =1,\ldots ,26)$ are Grassmann even
and the ghost fields $C$ and $\bar{C}$ are Grassmann odd.
The metric of the $OSp(26|2)$ vector space
is
\begin{equation}
\eta_{MN}
=
\begin{array}{c}
\\
\\
\\
\\
\mbox{\scriptsize $C$}\\
\mbox{\scriptsize $\bar{C}$}\\
\end{array}
\begin{array}{r}
\mbox{\scriptsize$C$}~~\mbox{\scriptsize$\bar{C}$}\hspace{6mm}\\
\left(
\begin{array}{ccc|cc}
 &             & & &  \\
 &\delta_{\mu\nu}& & &  \\
 &             & & &  \\\hline
 &             & &0&-i\\
 &             & &i&0 \\
\end{array}
\right)
\end{array}
 =\eta^{MN}~,
\end{equation}
where we have taken the Euclidean signature for the physical 
space-time.
$X^M$ are Fourier expanded in the usual way and 
we obtain the non-zero oscillation modes 
\begin{eqnarray}
\alpha^{M}_{n}
&=&
\left(\alpha^{\mu}_{n},-\gamma_{n},\bar{\gamma}_{n}\right)~,
\nonumber
\\
\tilde{\alpha}^{M}_{n}
&=&
\left(\tilde{\alpha}^{\mu}_{n},-\tilde{\gamma}_{n},
      \tilde{\bar{\gamma}}_{n}\right)\quad (n\neq 0)~,
\end{eqnarray}
and the zero modes 
\begin{eqnarray}
& &
x^{M}
=
\left(x^{\mu},C_{0},\bar{C}_{0}\right)~,
\nonumber
\\
& &
\alpha^{M}_{0}=\tilde{\alpha}^{M}_{0}
=p^M
=\left(p^{\mu},-\pi_{0},\bar{\pi}_{0}
   \right)~.
\end{eqnarray}
They satisfy the canonical commutation relations
\begin{equation}
[x^{N},p^{M}\}=i\eta^{NM}~,
\quad [\alpha^{N}_{n},\alpha^{M}_{m} \}=n\eta^{NM}\delta_{n+m,0}~,
\quad [\tilde{\alpha}^{N}_{n},\tilde{\alpha}^{M}_{m} \}
  = n \eta^{NM} \delta_{n+m,0}
\end{equation}
for $n,m\neq 0$,
where the graded commutator $[A,B\}$ denotes the anti-commutator
when $A$ and $B$ are both fermionic operators and
the commutator otherwise.

We define the Fock vacuum $|0\rangle$ in the usual way and 
take the momentum representation for the wave functions for the zero modes. 
The integration measure for the zero-modes 
of the $r$-th string is defined as 
\begin{equation}
dr
\equiv
 \frac{\alpha_rd\alpha_r}{2}
 \frac{d^{26}p_r}{(2\pi)^{26}} \, 
i  d\bar{\pi}_0^{(r)} d\pi_0^{(r)}~.
 \label{eq:zeromodemeasure}
\end{equation}
It is convenient to define the measure $d'r$
for $p^{\mu}_r,\pi^{(r)}_{0},\bar{\pi}^{(r)}_{0}$ as
\begin{equation}
d'r = \frac{d^{26} p_{r}}{(2 \pi)^{26}}
      i d\bar{\pi}^{(r)}_{0} d\pi^{(r)}_{0}~.
\label{eq:dprimer}
\end{equation}

\subsubsection*{action}
The action of the $OSp$ invariant string field theory takes the form
\begin{eqnarray}
\lefteqn{
S = \int dt  \left[
  \frac{1}{2}
   \int d1d2\, \left\langle R(1,2) \left|\Phi\right\rangle_{1}
               \right.
       \left( i\frac{\partial}{\partial t}
                  -\frac{L_0^{(2)}+\tilde{L}_0^{(2)}-2}{\alpha_2}
       \right)  \left|\Phi\right\rangle_{2}
   \right.} \hspace{3.5em}\nonumber\\
&& \left.
{} + \frac{2g}{3}
\int d1d2d3 \, \left\langle V_3^0(1,2,3)\right|
\Phi\rangle_1|\Phi\rangle_2|\Phi\rangle_3
\right]~.
\label{eq:actionOSp}
\end{eqnarray}
Here $\left\langle R(1,2)\right|$ is the reflector given as
\begin{equation}
\left\langle R(1,2)\right|
= \delta (1,2)
\; {}_{12}\!\langle 0|
 \, e^{E(1,2)}\, \frac{1}{\alpha_1}~,
 \label{eq:reflector}
\end{equation}
where
\begin{eqnarray}
{}_{12}\! \langle 0| 
&=& {}_{1}\!\langle 0| {}_{2}\!\langle 0|~,
 \nonumber\\
E(1,2)
 &=&
  -\sum_{n=1}^\infty\frac{1}{n}
        \left(\alpha_{n}^{N(1)} \alpha_{n}^{M(2)}
        +\tilde{\alpha}_{n}^{N (1)} \tilde{\alpha}_{n}^{M(2)}
        \right)\eta_{NM}~,
\nonumber\\
\delta (1,2)
&=&
2 \delta (\alpha_1+\alpha_2)
(2\pi )^{26}\delta^{26}(p_1+p_2)
i (\bar{\pi}_0^{(1)}+\bar{\pi}_0^{(2)})
(\pi_0^{(1)}+\pi_0^{(2)})~.
\label{eq:reflector1.5}
\end{eqnarray}
$\langle V_{3}^0(1,2,3)|$ is the three-string vertex
given as
\begin{equation}
\left\langle V_{3}^{0}(1,2,3) \right|
 \equiv  \delta (1,2,3)
   \; {}_{123}\!\langle 0|e^{E (1,2,3)}
    \mathcal{P}_{123}
    \frac{|\mu (1,2,3)|^2}{\alpha_1\alpha_2\alpha_3}~,
\label{eq:V30}
\end{equation}
where
\begin{eqnarray}
{}_{123}\!\langle 0|
 &=& {}_{1}\!\langle 0 |\,{}_{2}\!\langle 0|\,{}_{3}\!\langle 0|~,
\nonumber\\
\mathcal{P}_{123}&=&\mathcal{P}_{1}\mathcal{P}_{2}\mathcal{P}_{3}~,
\quad \mathcal{P}_{r}=\int^{2\pi}_{0} \frac{d\theta}{2\pi}
      e^{i\theta \left(L^{(r)}_{0}-\tilde{L}^{(r)}_{0}\right)}~,
\nonumber\\
\delta (1,2,3)
&=&
2 \delta \Bigl( \sum_{s=1}^3 \alpha_s \Bigr)
 (2\pi )^{26} \delta^{26} \Bigl( \sum_{r=1}^3 p_r \Bigr)
\, i\Bigl(\sum_{r^\prime =1}^3\bar{\pi}_0^{(r^\prime )} \Bigr)
\Bigl( \sum_{s^\prime =1}^3\pi_0^{(s^\prime )} \Bigr)~,
\nonumber\\
E(1,2,3)
&=&
\frac{1}{2}\sum_{n,m\geq 0}\sum_{r,s=1}^{3}
\bar{N}_{nm}^{rs}
 \left( \alpha_n^{N (r)} \alpha_{m}^{M (s)}
        +\tilde{\alpha}_n^{N (r)} \tilde{\alpha}_{m}^{M (s)}
  \right)\eta_{NM}~,
\nonumber \\
\mu(1,2,3) & = &
  \exp\left(-\hat{\tau}_{0} \sum_{r=1}^{3}\frac{1}{\alpha_{r}} 
      \right)~,
    \quad \hat{\tau}_{0}
    =\sum_{r=1}^{3} \alpha_{r} \ln \left|\alpha_{r}\right|~.
\label{eq:V3variables}
\end{eqnarray}
Here $\bar{N}^{rs}_{nm}$ denote the Neumann coefficients
associated with the joining-splitting type of three-string
interaction~\cite{Kaku:1974zz}%
\cite{Mandelstam:1973jk}\cite{Cremmer:1974ej}.
$g$ is the coupling constant for strings. 
In this paper, we take $g>0$. 

The string field $\Phi$ is taken to be Grassmann even
and subject to the level matching condition
$\mathcal{P}\Phi=\Phi$ and the
reality condition
\begin{equation}
\langle \Phi_{\mathrm{hc}} |
  = \langle \Phi |~.
  \label{eq:reality1}
\end{equation}
Here
$\langle \Phi_{\mathrm{hc}} |
\equiv \left( | \Phi \rangle \right)^{\dagger}$
denotes the hermitian conjugate of
$| \Phi \rangle$,
and $\langle \Phi |$ denotes the BPZ conjugate
of $|\Phi \rangle$ defined as 
\begin{equation}
{}_{2} \! \langle \Phi |
=
\int d1\, \langle R(1,2)|\Phi \rangle_1~.
\label{eq:BPZ}
\end{equation}
We also define
\begin{equation}
|R(1,2)\rangle
\equiv
\delta (1,2)\frac{1}{\alpha_1}e^{E^\dagger (1,2)}|0\rangle_{12}~,
\end{equation}
so that 
\begin{equation}
\int d1{}_1\langle \Phi |R(1,2)\rangle
=
|\Phi \rangle_2~. 
\end{equation}

\subsubsection*{BRST transformation}
The action (\ref{eq:actionOSp}) is
invariant under the BRST transformation
\begin{equation}
\delta_{\mathrm{B}} \Phi
 =Q_{\mathrm{B}} \Phi +g\Phi *\Phi~.
\label{BRST}
\end{equation}

The BRST operator $Q_{\mathrm{B}}$ is
defined~\cite{Siegel:1986zi}\cite{Bengtsson:1986yj} as 
\begin{eqnarray}
 Q_{\mathrm{B}}
 &=& \frac{C_0}{2\alpha}(L_0+\tilde{L}_{0} - 2)
   -i\pi_0\frac{\partial}{\partial \alpha}
  \nonumber\\
&& 
    {} +\frac{i}{\alpha}\sum_{n=1}^{\infty}
      \left(
        \frac{\gamma_{-n}L_n-L_{-n}\gamma_n}{n}
          +\frac{\tilde{\gamma}_{-n}\tilde{L}_n
                 -\tilde{L}_{-n}\tilde{\gamma}_n}{n}\right)~.
  \label{eq:brst-charge}
\end{eqnarray}
Here $L_{n}$ and $\tilde{L}_{n}$ $(n\in\mathbb{Z})$ are
the Virasoro generators given by
\begin{equation}
L_n = \frac{1}{2} \sum_{m\in \mathbb{Z}}
     \maru \alpha_{n+m}^{N}\alpha_{-m}^{M} \eta_{NM}\maru~,
\quad
\tilde{L}_n = \frac{1}{2} \sum_{m\in\mathbb{Z}}
      \maru \tilde{\alpha}_{n+m}^{N}
              \tilde{\alpha}_{-m}^{M} \eta_{NM}
         \maru~,
\end{equation}
where the symbol $\maru\ \maru$ denotes
the normal ordering of the oscillators
in which the non-negative modes should be placed to the right
of the negative modes. 

The $\ast$-product $\Phi \ast \Psi$
of two arbitrary closed string fields
$\Phi$ and $\Psi$ is given as
\begin{equation}
\left|\Phi *\Psi \right\rangle_4
=
\int d1d2d3\, \left\langle V_3(1,2,3)\left|
     \Phi\right\rangle_1 \right.
     \left|\Psi \right\rangle_2
     \left| R(3,4) \right\rangle~,
 \label{eq:star-prod}
\end{equation}
where 
\begin{eqnarray}
\left\langle V_3(1,2,3)\right|
=
\delta (1,2,3)
   \; {}_{123}\!\langle 0|e^{E (1,2,3)}
    C(\rho_I)
    \mathcal{P}_{123}
    \frac{|\mu (1,2,3)|^2}{\alpha_1\alpha_2\alpha_3}~.
\label{eq:V3}
\end{eqnarray}
$\rho_{I}$ denotes the interaction point.

\subsubsection*{canonical quantization}
Since the action (\ref{eq:actionOSp}) and the formulation of the 
$OSp$ invariant string field theory are quite similar to those of the 
light-cone gauge string field theory, we can perform the canonical quantization 
in an analogous way. We can decompose the string field as 
\begin{equation}
\left| \Phi \right\rangle
         = \left| \psi  \right\rangle
            + \left| \bar{\psi}  \right\rangle~,
\end{equation}
where $|\psi \rangle$ is the part with positive $\alpha$ and 
$|\bar{\psi}\rangle$ is the one with negative $\alpha$. 
{}From the kinetic term of eq.(\ref{eq:actionOSp}), we can see that 
they satisfy the canonical commutation relation 
\begin{equation}
  \left[ \left|\psi\right\rangle_{r},
       \left|\bar{\psi} \right\rangle_{s} \right]
  =\left| R (r,s) \right\rangle~.
\label{eq:ccr}
\end{equation}
{}From the hermiticity defined in eq.(\ref{eq:reality1}), 
one can deduce that $\langle \psi |$ and $\langle \bar{\psi}|$ 
are hermitian conjugate to $| \bar{\psi} \rangle$ and
$| \psi \rangle$, respectively. 
We identify $|\psi \rangle$ with the annihilation mode and 
$|\bar{\psi}\rangle$ with the creation mode. 
Accordingly we define the vacuum state
$|0\rangle\!\rangle$ in the
second quantization as
\begin{equation}
|\psi \rangle |0\rangle\!\rangle =0~,
\quad
\langle\!\langle 0 | \langle \bar{\psi} | =0~.
\end{equation}

\section{Overlap of Three-String Vertex 
         with One Boundary State}\label{sec:tree}

In this appendix, we evaluate the string vertex
$\langle V_{2}(1,2);T|$ introduced in eq.(\ref{eq:treevertex})
for $T=\epsilon$.
$\langle V_{2}(1,2);T|$ can be expressed as
\begin{equation}
\langle V_{2}(1,2);T|
=
\langle V_2^{0}(1,2);T|
C\left( \rho_I \right)
\mathcal{P}_{12}
~,
\end{equation}
where
\begin{equation}
\langle V_2^{0}(1,2);T| = \int d'3 
\delta (1,2,3)
   \; {}_{123}\!\langle 0|e^{E (1,2,3)}|B_{0}\rangle_3^T 
    \frac{|\mu (1,2,3)|^2}{\alpha_1\alpha_2\alpha_3}~.
\label{eq:v20t}
\end{equation} 
Here we present the calculations for the case $\alpha_1,\alpha_2 <0$.

\subsubsection*{Mandelstam mapping}

The vertex $\langle V_2^{0}(1,2);T|$ is proportional to
the one that is determined by the prescription of LeClair,
Peskin and Preitschopf (LPP)~\cite{LeClair:1988sp}.
We refer to the latter as the LPP vertex.
As we will see, we can calculate it
by using the Mandelstam mapping which maps 
the upper half plane to the worldsheet in Fig.\ref{fig:uhp1}.

\begin{figure}[htbp]
\begin{minipage}{19em}
\begin{center}
	\includegraphics[width=18.5em]{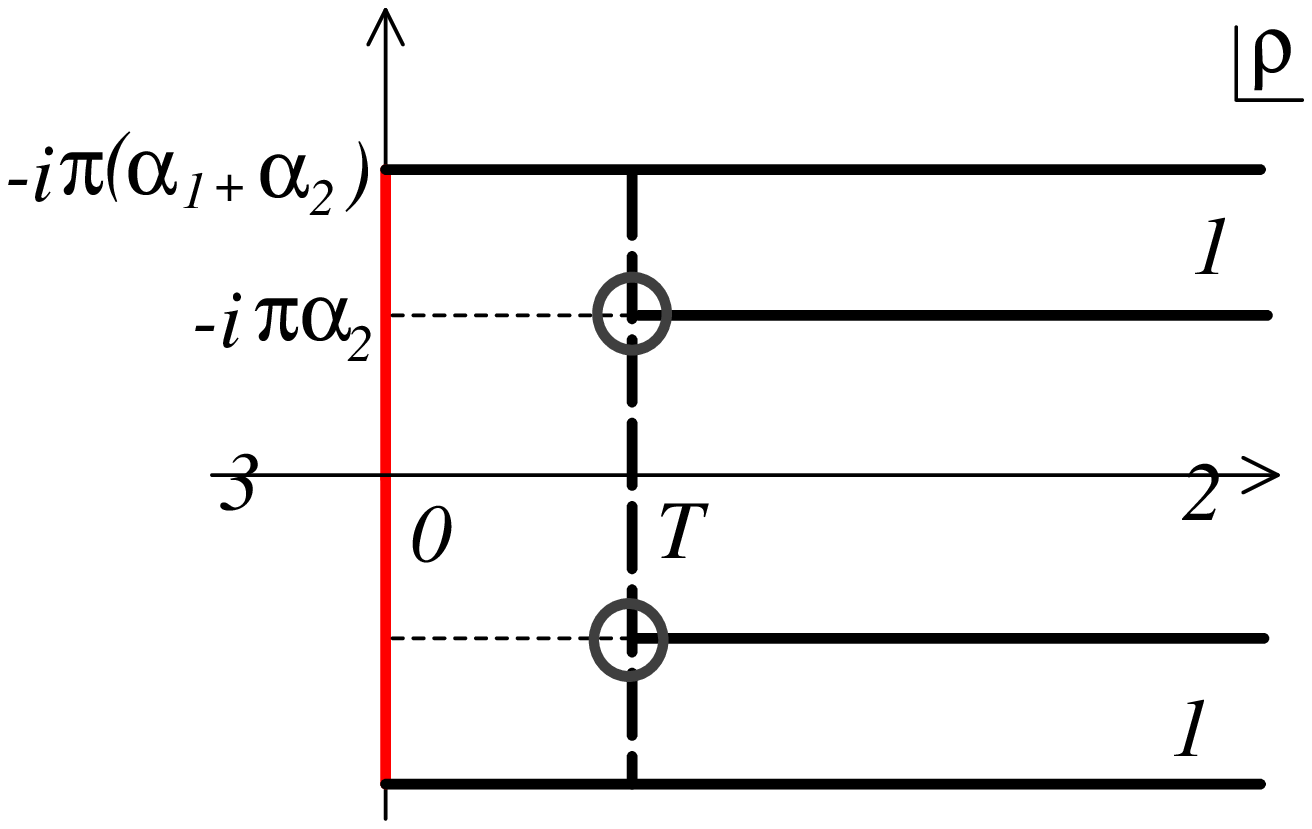}
	\caption{The $\rho$-plane corresponding to the
	         string diagram depicted in Fig.\ref{fig:uhp1}.}
	\label{fig:rhotree}
\end{center}
\end{minipage}
\hfill
\begin{minipage}{19em}
\begin{center}
	\includegraphics[width=18em]{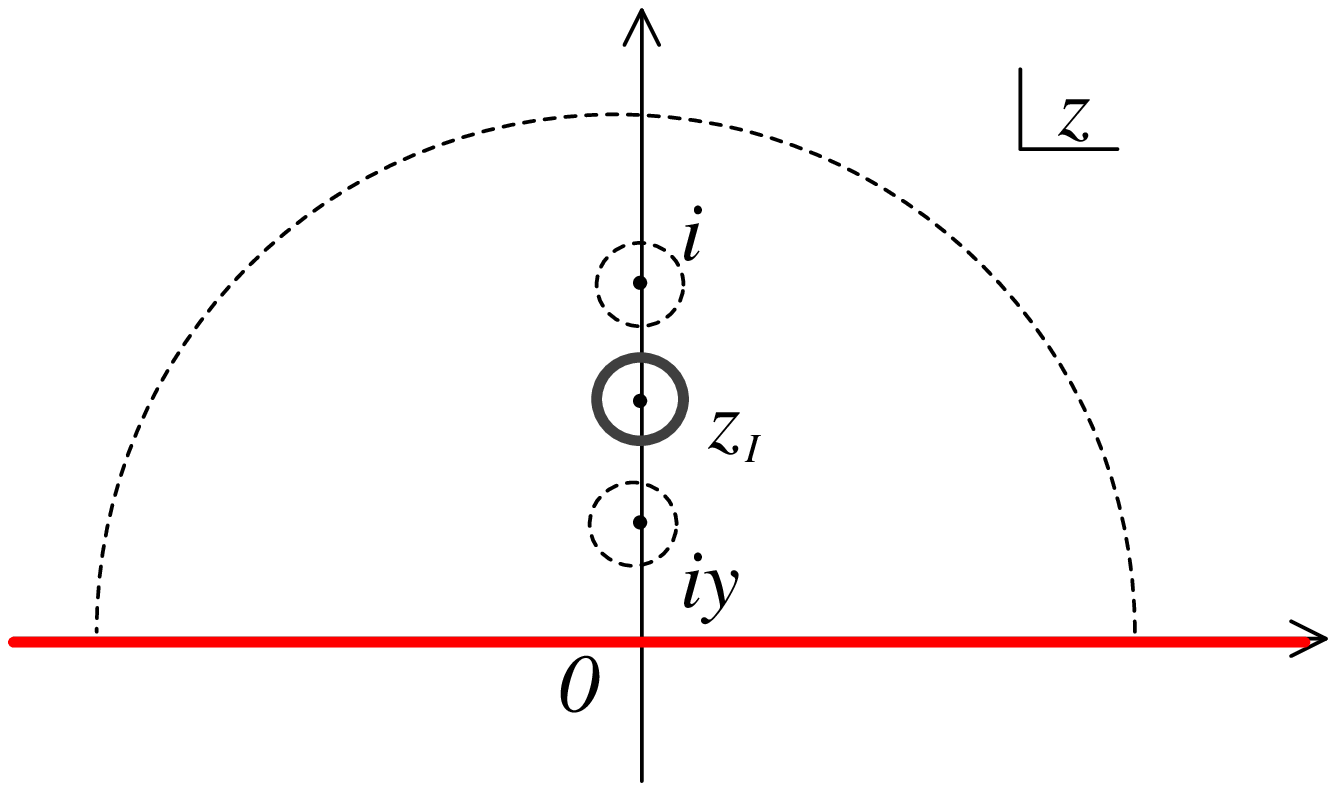}
	\caption{The upper half $z$-plane.}
	\label{fig:uhp2}
\end{center}
\end{minipage}
\end{figure}

Let us introduce a complex coordinate $\rho$ on the worldsheet so that
the string diagram in Fig.\ref{fig:uhp1} can be identified with
the region depicted in Fig.\ref{fig:rhotree} on the $\rho$-plane.
Each portion of the $\rho$-plane corresponding to the $r$-th
external string $(r=1,2)$ is identified with
the unit disk $|w_{r}| \leq 1$ of string $r$ by the relation
\begin{eqnarray}
&& \rho = \alpha_{r} \zeta_{r} + T + i\beta_{r}~,
\quad
\beta_{r} = - \alpha_{2} \pi - \alpha_{r} \sigma^{(r)}_{I}~,
\nonumber\\
&&\zeta_r (=\tau_r +i \sigma_r) = \ln w_r~,
\qquad \tau_{r} \leq 0~,
\quad - \pi \leq \sigma_{r} \leq \pi~.
\label{eq:unitdisk}
\end{eqnarray}
Here $\rho_{I}= T-i \pi \alpha_{2}$ is the interaction point
on the $\rho$-plane and  $\sigma_{I}^{(r)}$ is the value of
the $\sigma_r$ coordinate
where the $r$-th string interacts.
We set $\sigma_{I}^{(1)} = \pi$ and $\sigma_{I}^{(2)}=-\pi$.
Therefore we have
\begin{equation}
\beta_{1} = -(\alpha_{1} + \alpha_{2}) \pi~,
\quad
\beta_{2} = 0~.
\end{equation}

The string diagram described by Fig.\ref{fig:uhp1} has one hole
and two punctures at infinity corresponding to the two external strings,
strings $1$ and $2$. Since the topology of this diagram is a disk
with two punctures, 
the $\rho$-plane~(Fig.\ref{fig:rhotree}) can be
mapped to the complex upper half $z$-plane~(Fig.\ref{fig:uhp2})
with two punctures.
These two surfaces are related by the Mandelstam mapping
\begin{equation}
\rho (z) = \alpha_{1} \ln \frac{z - Z_{1}}{z - \bar{Z}_{1}}
     + \alpha_{2} \ln \frac{z - Z_{2}}{z - \bar{Z}_{2}}~,
\label{eq:uhp}
\end{equation}
where 
the point $z=Z_{r}$ $(r=1,2)$ is the puncture corresponding to
the origin of the unit disk $|w_{r}|<1$ of string $r$.
We can set $Z_{1}=iy$ and $Z_{2}=i$, where
$y$ is a real parameter with $0<y<1$.
The interaction point $z_{I}$ on the $z$-plane is determined
by $\frac{d\rho}{dz}(z_{I}) = 0$. This yields
\begin{equation}
z_{I} = 
i \sqrt{\frac{(\alpha_{1} + \alpha_{2}y)y}
             {\alpha_{1}y + \alpha_{2}}}~.
\label{eq:intpt-tree}
\end{equation}
Here we have used $\alpha_{1},\alpha_{2} <0$,
$0<y<1$ and $\mathrm{Im}\, z_{I} > 0$.
Eq.(\ref{eq:intpt-tree}) leads to
\begin{equation}
T = \mathrm{Re}\, \rho(z_{I})
 = \alpha_{1} \ln \left|\frac{z_{I} - iy}{z_{I} + iy} \right|
   + \alpha_{2} \ln \left| \frac{z_{I} -i}{z_{I} +i} \right|~.
 \label{eq:tree-T}
\end{equation}
{}From this relation, we find that in the small $T$ limit, 
$T = \epsilon \ll 1$,
we have
\begin{equation}
y \simeq \frac{1}{16 \alpha_{1} \alpha_{2}} \epsilon^{2} +
\mathcal{O}(\epsilon^{4})~.
\label{eq:smallT-y-tree}
\end{equation}
For later use, we consider the limit $T\rightarrow \infty$ as well.
In this limit, $y\sim 1$. In fact,
\begin{equation}
T \simeq \hat{\tau}_{0} - (\alpha_{1}+\alpha_{2}) \ln 2
 + (\alpha_{1} + \alpha_{2}) \ln (1-y)
 + \mathcal{O}(1-y)~,
\label{eq:largeT-y-tree}
\end{equation}
where
\begin{equation}
\hat{\tau}_{0} = \alpha_{1} \ln |\alpha_{1}|
+\alpha_{2} \ln |\alpha_{2}|
 - (\alpha_{1} + \alpha_{2}) \ln |\alpha_{1}+\alpha_{2}|~.
\label{eq:tau0}
\end{equation}

\subsubsection*{Neumann coefficients}
The real axis of the $z$-plane corresponds to the worldsheet
boundary attached to $|B_{0}\rangle_{3}$.
Because of the boundary conditions (\ref{eq:boundarycond})
satisfied by the worldsheet
variables $X^{N}=(X^{\mu}, X^{i}, C,\bar{C})$
on the boundary state $|B_{0}\rangle$,
the two-point functions of $X^{N}(z,\bar{z})$ on the $z$-plane
become
\begin{equation}
G^{NM}_{\mathrm{UHP}} (z,\bar{z};z',\bar{z}')
 = \langle X^{N}(z,\bar{z}) X^{M}(z',\bar{z}')
   \rangle
= -\eta^{NM} \ln |z-z'|^{2} - D^{NM} \ln |z-\bar{z}'|^{2}~,
\label{eq:two-pt-tree}
\end{equation}
where $D^{NM}$ is the tensor introduced in eq.(\ref{eq:metricD}).

The vertex $\langle V_{2}^{0}(1,2);T|$ introduced in eq.(\ref{eq:v20t})
takes the form
\begin{equation}
\langle V_{2}^{0} (1,2);T| =
2 \delta (\alpha_{1}+\alpha_{2}+\alpha_{3})
(2\pi)^{p+1} \delta^{p+1}_{\mathrm{N}} (p_{1}+p_{2})
\, \mathcal{K}_{2}(1,2;T)
\, \langle V_{2,\mathrm{LPP}}^{0}(1,2);T|~,
\end{equation}
where $\langle V_{2,\mathrm{LPP}}^{0}(1,2);T|$
is the LPP vertex, and the factor $\mathcal{K}_{2}(1,2;T)$
depends only on the zero-modes and the moduli.
The LPP vertex has the structure
\begin{eqnarray}
\lefteqn{
\langle V_{2,\mathrm{LPP}}^{0}(1,2);T|
} \nonumber\\
 && =
  {}_{12} \langle 0| \exp \left[ \sum_{n,m=0}^{\infty}
   \sum_{r,s=1,2}  \Bigl\{ \frac{1}{2} \left( 
    \bar{N}^{(2)rs}_{\ \ nm}\,
           \alpha^{N(r)}_{n} \, \alpha^{M(s)}_{m}
   + \bar{N}^{(2)\tilde{r}\tilde{s}}_{\ \ nm} \,
           \tilde{\alpha}^{N(r)}_{n} \tilde{\alpha}^{M(s)}_{m}
   \right)\eta_{NM} \right. 
 \nonumber\\
 && \hspace{9.5em} \left. + \frac{1}{2} \left( 
   \bar{N}^{(2)r\tilde{s}}_{\ \ nm} \, 
        \alpha^{N(r)}_{n} \tilde{\alpha}^{M(s)}_{m}
        +\bar{N}^{(2)\tilde{r}s}_{\ \ nm} \,
         \tilde{\alpha}^{N(r)}_{n} \alpha^{M(s)}_{m}
         \right) D_{NM}
   \Bigr\}
   \right],
\end{eqnarray}
for Wick's theorem to hold. 
The Neumann coefficients $\bar{N}^{(2) rs}_{\ \ nm}$,
$\bar{N}^{(2)\tilde{r}\tilde{s}}_{\ \ nm}$ and
$\bar{N}^{(2)r\tilde{s}}_{\ \ nm}$ 
are determined by requiring
that the following equation should hold~\cite{LeClair:1988sp},
\begin{eqnarray}
&&
\int d'1 d'2 \,
\langle V_{2,\mathrm{LPP}}^{0}(1,2);T|
 X^{N (r)} (w_{r},\bar{w}_{r}) X^{M(s)}(w'_{s},\bar{w}'_{s})
 |0\rangle_{12}
 \nonumber\\
&& \hspace{10.5em}
\times 
 \prod_{r'=1,2}
  (2\pi)^{26}\delta^{26}(p_{r'})
            i \bar{\pi}_{0}^{(r')}  \pi_{0}^{(r')}
 = G^{NM}_{\mathrm{UHP}} (z_{r},\bar{z}_{r};z'_{s},\bar{z}'_{s})~,~~~~
\label{eq:LPP}
\end{eqnarray}
where $z_{r}$ and $z'_{s}$ are the points on the $z$-plane
corresponding to the points $w_{r}$ and $w'_{s}$ on the unit
disks of strings $r$ and $s$, respectively.

Using eq.(\ref{eq:LPP}), one can show that the Neumann coefficients 
are given as
\begin{eqnarray}
& &
\bar{N}^{(2) rs}_{\ \ nm}=
\left(
\bar{N}^{(2) \tilde{r}\tilde{s}}_{\ \ nm}
\right)^\ast
=
\frac{1}{nm}
\oint_{Z_r}\frac{dz}{2\pi i}
\oint_{Z_s}\frac{dz^\prime}{2\pi i}
\frac{e^{-n\zeta_r(z)-m\zeta^\prime_s (z^\prime )}}{(z-z^\prime )^2}~,
\nonumber
\\
& &
\bar{N}^{(2) r\tilde{s}}_{\ \ nm}=
\left(
\bar{N}^{(2) \tilde{r}s}_{\ \ nm}
\right)^\ast
=
\frac{1}{nm}
\oint_{Z_r}\frac{dz}{2\pi i}
\oint_{\bar{Z}_s}\frac{d\bar{z}^\prime}{2\pi i}
\frac{e^{-n\zeta_r(z)-m\bar{\zeta}^\prime_s (\bar{z}^\prime )}}{(z-\bar{z}^\prime )^2}~,
\nonumber
\\
& &
\bar{N}^{(2) rs}_{\ \ n0}=
\left(
\bar{N}^{(2) \tilde{r}\tilde{s}}_{\ \ n0}
\right)^\ast
=
\frac{1}{n}\oint_{Z_r}\frac{dz}{2\pi i}
\frac{e^{-n\zeta_r(z)}}{z-Z_s}~,
\nonumber
\\
& &
\bar{N}^{(2) r\tilde{s}}_{\ \ n0}=
\left(
\bar{N}^{(2) \tilde{r}s}_{\ \ n0}
\right)^\ast
=
\frac{1}{n}\oint_{Z_r}\frac{dz}{2\pi i}
\frac{e^{-n\zeta_r(z)}}{z-\bar{Z}_s}~,
\nonumber
\\
& &
\bar{N}^{(2) rs}_{\ \ 00}
 =\left( \bar{N}^{(2) \tilde{r}\tilde{s}}_{\ \ 00}\right)^{\ast}
=
\ln (Z_r-Z_s)
~~~(r\neq s)~,
\nonumber
\\
& &
\bar{N}^{(2) r\tilde{s}}_{\ \ 00}
 =\left( \bar{N}^{(2) \tilde{r}s}_{\ \ 00} \right)^{\ast}
=
\ln (Z_r-\bar{Z}_s)
~~~(r\neq s)~,
\nonumber
\\
& &
\bar{N}^{(2) rr}_{\ \ 00}
 =\left( \bar{N}^{(2) \tilde{r}\tilde{r}}_{\ \ 00} \right)^{\ast}
=
\ln (Z_r-\bar{Z}_r)
-\sum_{s\neq r}\frac{\alpha_s}{\alpha_r}
\left\{
\ln (Z_r-Z_s)-\ln (Z_r-\bar{Z}_s)
\right\}
+\frac{T+i\beta_{r}}{\alpha_r}~,
\nonumber
\\
& &
\bar{N}^{(2) r\tilde{r}}_{\ \ 00}
 =\left(\bar{N}^{(2) \tilde{r}r}_{\ \ 00}\right)^{\ast}
=
\ln (Z_r-\bar{Z}_r)~,
\end{eqnarray}
for $n,m \geq 1$. 
Here we have used the convention for the orientation of
the $\bar{z}$ integration
such that $\oint_{0} \frac{d\bar{z}}{2\pi i} \frac{1}{\bar{z}} = 1$.

\subsubsection*{$\mathcal{K}_{2}(1,2;T)$}
The central charge of the worldsheet CFT
 of the $OSp$ invariant string theory is $24$ and not $0$.
Therefore
the Generalized Gluing and Resmoothing Theorem~\cite{LeClair:1988sj}
does not hold in this
case and thus $\mathcal{K}_2(1,2;T) \neq 1$.
Since the three-string vertex $\langle V_{3}^{0}(1,2,3)|$ 
is defined assuming that the $\rho$-plane is endowed with the metric
\begin{equation}
ds^2=d\rho d\bar{\rho}~,
\label{eq:rhometric}
\end{equation}
the oscillator independent part
$\mathcal{K}_2(1,2;T)$ is the partition function
of the CFT on the $\rho$-plane~(Fig.\ref{fig:rhotree}) 
with the metric given in eq.(\ref{eq:rhometric}).
As explained in \cite{Mandelstam:1985ww},  its dependence on $\alpha_{1}$,
$\alpha_{2}$ and the moduli $T$ can be determined through CFT technique
by evaluating the Liouville action
associated with the conformal mapping (\ref{eq:uhp})
between the $\rho$-plane and the 
upper half $z$-plane
with small circles
around $z_{I}$, 
$Z_r~(r=1,2)$ and $\infty$ excised.
Collecting the contributions from these holes,
we obtain
\begin{equation}
\mathcal{K}_2(1,2;T)
\propto
\left| \lim_{z\rightarrow\infty}
       \left(   z^2 \frac{d\rho (z)}{dz}  \right)
\right|^2
\prod_{r=1,2}
\left( |\alpha_r|^{-2} 
       \left| \frac{d w_r}{d z} 
               \left( Z_r \right)
       \right|^2
\right)
\left| \frac{d^{2}\rho}{dz^{2}}(z_I) 
          \right|^{-1}~.
\end{equation}
Therefore we can see that $\mathcal{K}_2(1,2;T)$ is expressed as
\begin{equation}
\mathcal{K}_2(1,2;T)
 = \mathcal{K}_{0}
   \frac{1}{\alpha_{1} \alpha_{2}}
   \sqrt{\frac{(\alpha_{1} y + \alpha_{2}) y}{\alpha_{1} + \alpha_{2} y}}
   \,
   \frac{(1-y^{2})^2}{(\alpha_{1} y + \alpha_{2})16 y^{2}}
   \;
    e^{-2\left(\frac{1}{\alpha_{1}}+\frac{1}{\alpha_{2}}\right) T
      -2 \left(\frac{\alpha_{2}}{\alpha_{1}} + \frac{\alpha_{1}}{\alpha_{2}}
       \right) \ln\frac{1+y}{1-y}}~,
\end{equation}
where $\mathcal{K}_{0}$ is a constant independent of
$\alpha_{1}$, $\alpha_{2}$ and $T$.
$\mathcal{K}_{0}$ can be determined by comparing the left and right
hand sides of the equation
\begin{eqnarray}
&&
\int d'1 d'2 d'3 \, \langle V^{0}_{3}(1,2,3)|B_{0}\rangle^{T}_{3}
 |0\rangle_{12} \prod_{r=1,2} (2 \pi)^{26} \delta^{26}(p_{r})
  i\bar{\pi}^{(r)}_{0} \pi^{(r)}_{0}
  \nonumber \\
&& \qquad
 = \int d'1 d'2 \;
 2\delta(\alpha_{1} + \alpha_{2} + \alpha_{3} )
 (2\pi)^{p+1} \delta^{p+1}_{\mathrm{N}}(p_{1}+p_{2}) \;
  \mathcal{K}_{2}(1,2;T)
 \nonumber\\
&& \hspace{7.5em}
  \times
  \langle V^{0}_{2,\mathrm{LPP}}(1,2);T| 0\rangle_{12}
  \prod_{r=1,2} (2 \pi)^{26} \delta^{26}(p_{r})
  i\bar{\pi}^{(r)}_{0} \pi^{(r)}_{0}~,
\end{eqnarray}
in the $T \rightarrow \infty$ limit.
One can readily evaluate the left hand side of the above equation
because the non-zero oscillation modes do not contribute
in this limit.
Using eq.(\ref{eq:largeT-y-tree}),
we find that $\mathcal{K}_{0}= - 1$.

\subsubsection*{$C(\rho_I)$}
The effect of inserting $C(\rho_I)$ can be described as follows:
\begin{eqnarray}
\langle V_{2,\mathrm{LPP}} (1,2);T|
&\equiv& \langle V^{0}_{2,\mathrm{LPP}}(1,2);T|C(\rho_{I})
 \nonumber\\
 &=&
 \langle V^{0}_{2,\mathrm{LPP}}(1,2);T|
 \left[ \sum_{n=0}^{\infty} \sum_{r=1,2}
   \left( M_{\mathrm{UHP}n}^{\ \ \ \ \; r} i \gamma^{(r)}_{n}
          + M^{\ \ \ \ \; \tilde{r}}_{\mathrm{UHP}n}
           i \tilde{\gamma}^{(r)}_{n} \right)
  \right]~.
\label{eq:Cint-tree}
\end{eqnarray}
The coefficients $M^{\ \ \ \ \; r}_{\mathrm{UHP}n}$ and
 $M^{\ \ \ \ \; \tilde{r}}_{\mathrm{UHP} n}$ can be
determined by the LPP prescription, i.e.\ we require that
\begin{eqnarray}
\lefteqn{
\int d'1 d'2
\langle V_{2,\mathrm{LPP}} (1,2);T| \bar{C}^{(r)}(w_{r},\bar{w}_{r})
|0\rangle_{12}   \prod_{r=1,2} (2\pi)^{26}\delta^{26}(p_{r})
                         i\bar{\pi}^{(r)}_{0} \pi^{(r)}_{0}
} \nonumber\\
&& =
  G^{C\bar{C}}_{\mathrm{UHP}} (z_{I},\bar{z}_{I};z_{r},\bar{z}_{r})
= i \Bigl[ \ln (z_{I} - z_{r}) + \ln (\bar{z}_{I} - \bar{z}_{r})
              - \ln (z_{I} - \bar{z}_{r}) -\ln (\bar{z}_{I} - z_{r})
       \Bigr].~~~~~
\end{eqnarray}
This yields
\begin{equation}
M^{\ \ \ \ \; r}_{\mathrm{UHP}n}
 = (M^{\ \ \ \ \; \tilde{r}}_{\mathrm{UHP}n})^{\ast}
= -\frac{i}{n} \oint_{Z_{r}} \frac{dz_{r}}{2 \pi i}
   \, e^{-n\zeta_{r}(z_{r})}
 \left( \frac{i}{z_{r}-z_{I}} - \frac{i}{z_{r} -\bar{z}_{I}} \right)
\end{equation}
for $n \geq 1$ and
\begin{equation}
M^{\ \ \ \ \; r}_{\mathrm{UHP} 0}
 + M^{\ \ \ \ \; \tilde{r}}_{\mathrm{UHP} 0}
 = \ln (z_{I} - Z_{r}) + \ln (\bar{z}_{I} -\bar{Z}_{r})
   - \ln (z_{I} - \bar{Z}_{r}) - \ln (\bar{z}_{I} - Z_{r})~.
\end{equation}

\subsubsection*{$\langle V_{2}(1,2);\epsilon|$}
Collecting the results obtained in the above, we eventually
get the vertex $\langle V_{2}(1,2);T|$.
Now that we obtain the complete expression
of the vertex
$\langle V_{2}(1,2);T|$, let us consider the $T = \epsilon \to 0$ limit.
It is intuitively obvious that $\langle V^{0}_{2,\mathrm{LPP}}(1,2);\epsilon|$ 
is proportional to a product of boundary states in this limit.
It is straightforward to show that 
\begin{equation}
(2\pi)^{p+1} \delta^{p+1}_{\mathrm{N}} (p_{1}+p_{2})
\langle V^{0}_{2,\mathrm{LPP}}(1,2);\epsilon|
\sim
\frac{1}{(16\pi)^{\frac{p+1}{2}}(-\ln \epsilon )^{\frac{p+1}{2}}}
{}_{1}^{\epsilon} \langle B_{0}|
 \;  {}_{2}^{\epsilon} \langle B_{0}|~,
\label{eq:V0LPP}
\end{equation}
in the leading order. 
Therefore, in evaluating
$\langle V_{2}(1,2);\epsilon |
=
\langle V_2^{0}(1,2);\epsilon |
C\left( \rho_I \right)
\mathcal{P}_{12}$
, 
only the term proportional to $\pi_0^{(r)}$ from $C(\rho_I)$ 
survives the level matching condition and we obtain
\begin{eqnarray}
\lefteqn{
(2\pi)^{p+1} \delta^{p+1}_{\mathrm{N}} (p_{1}+p_{2})
\langle V_{2,\mathrm{LPP}} (1,2);\epsilon |\mathcal{P}_{12}
}\nonumber\\
&&
\sim
\frac{1}{(16\pi)^{\frac{p+1}{2}}(-\ln \epsilon )^{\frac{p+1}{2}}}
{}_{1}^{\epsilon} \langle B_{0}|
\;  {}_{2}^{\epsilon} \langle B_{0}|
\left(
\frac{i}{\alpha_1}\pi_0^{(1)}+ \frac{i}{\alpha_{2}} \pi^{(2)}_{0} 
\right)
\mathcal{P}_{12}~.
\label{eq:CrhoItree}
\end{eqnarray}
Evaluating $\mathcal{K}_2(1,2;\epsilon)$, we finally obtain
\begin{eqnarray}
\langle V_{2}(1,2);\epsilon| 
&\sim&
2\delta (\alpha_{1}+\alpha_{2}+\alpha_{3})
 \frac{1}{(16 \pi)^{\frac{p+1}{2}}}
 \frac{4}{\epsilon^{2} (- \ln \epsilon)^{\frac{p+1}{2}}}
\nonumber
\\
& &
\hspace{3cm}
\times
{}^{\epsilon}_{1}\langle B_{0}|
 \;  {}^{\epsilon}_{2}\langle B_{0}|
 \left( \frac{i}{\alpha_{1}} \pi^{(1)}_{0}
                  + \frac{i}{\alpha_{2}} \pi^{(2)}_{0} \right)
  \mathcal{P}_{12}
 ~.
\label{eq:v2(12)}
\end{eqnarray}

The case $\alpha_{1},\alpha_{2} >0$ can be treated in the same way and 
one can show that 
eq.(\ref{eq:v2(12)}) holds also in this case.

\section{Overlap of Three-String Vertex
         with Two Boundary States}\label{sec:torus}

In this appendix, we investigate the vertex
$\langle V_{1}(3);\epsilon|$ introduced in eq.(\ref{eq:loopvertex}).
The calculation proceeds in the same way as the one above. 
We begin by expressing $\langle V_{1}(3);T|$ as
\begin{equation}
\langle V_{1}(3);T|
=
\langle V_{1}^{0}(3);T|C(\rho_I)\mathcal{P}_3~,
\end{equation}
where
\begin{equation}
\langle V_{1}^{0}(3);T|
 \equiv \int d'1   d'2 \,
\delta (1,2,3)
   \; {}_{123}\!\langle 0|e^{E (1,2,3)}
   |B_{0}\rangle^{T}_{1} 
   |B_{0}\rangle^{T}_{2} 
\frac{|\mu (1,2,3)|^2}{\alpha_1\alpha_2\alpha_3}~.
\end{equation}
Here we present the calculations for the case $\alpha_{1},\alpha_{2} >0$. 

\subsubsection*{Mandelstam mapping}
\begin{figure}[htbp]
\begin{minipage}{18em}
\begin{center}
	\includegraphics[width=17.5em]{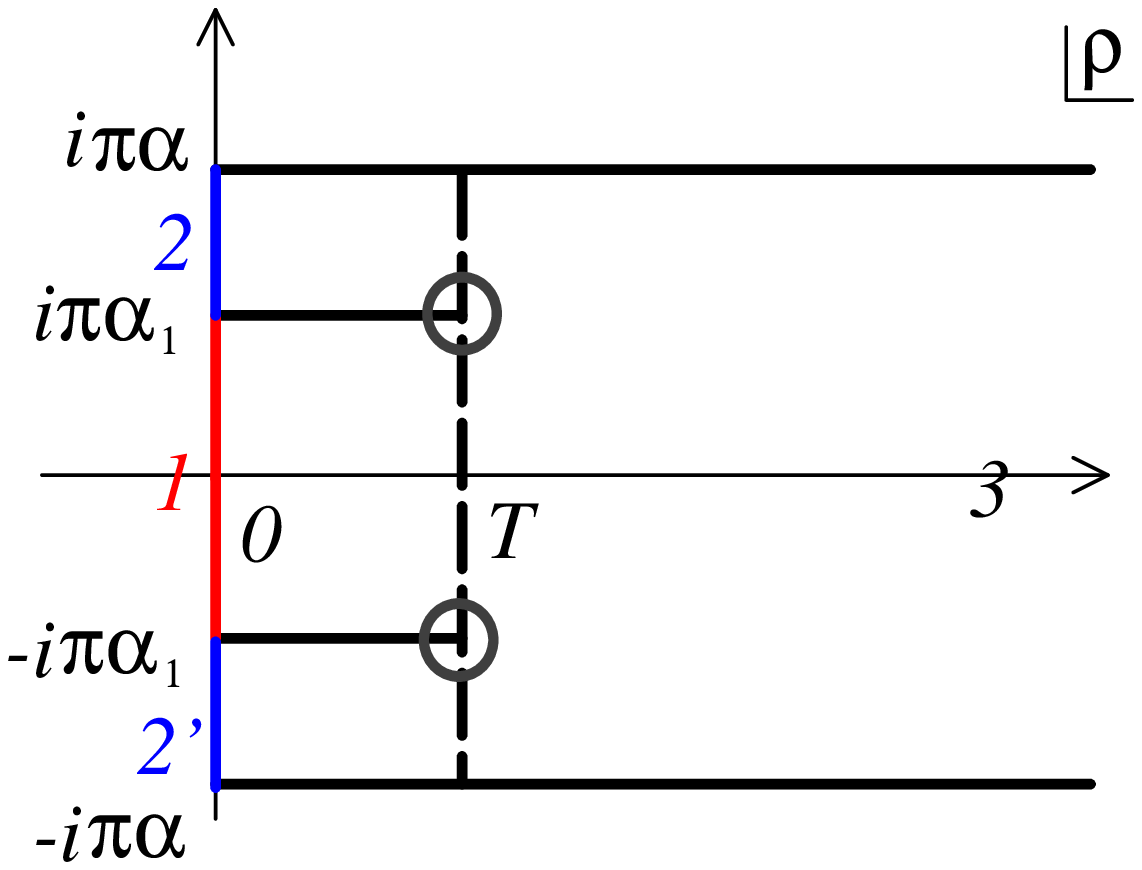}
	\caption{The $\rho$-plane corresponding to the string diagram
	         depicted by Fig.\ref{fig:annulus1}.}
	\label{fig:rhoannulus}
\end{center}
\end{minipage}
\hfill
\begin{minipage}{16.5em}
\begin{center}
	\includegraphics[width=16em]{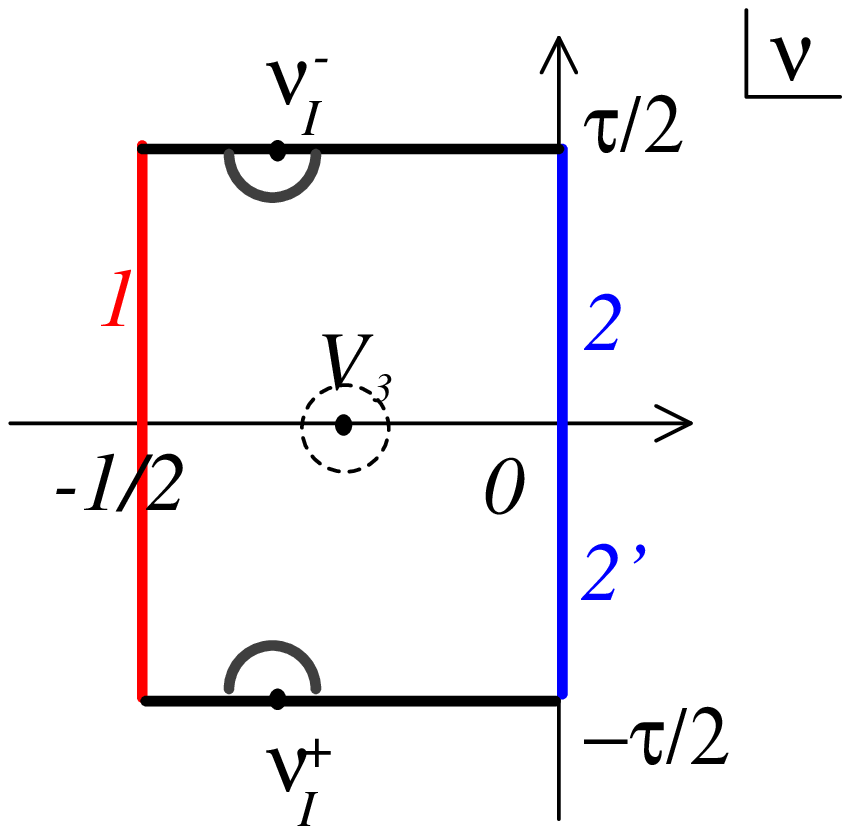}
	\caption{The rectangle on the $\nu$-plane.}
	\label{fig:rectangle}
\end{center}
\end{minipage}
\end{figure}

The complex $\rho$-plane indicating
the string diagram~Fig.\ref{fig:annulus1}
is described by Fig.\ref{fig:rhoannulus}.
The region of the $\rho$-plane corresponding to the external string,
string $3$,
is identified with the unit disk $|w_{3}|\leq 1$ of
this string through the relation
\begin{eqnarray}
&&\rho = \alpha_{3} \zeta_{3} + T + i\beta_{3}~,
\quad \beta_{3} = \alpha_{1} \pi - \alpha_{3} \sigma^{(3)}_{I}~,
\nonumber\\
&&\zeta_{3}(=\tau_{3} + i\sigma_{3}) = \ln w_{3}~,
\qquad
\tau_{3} \leq 0~, \quad -\pi \leq \sigma_{3} \leq \pi~.
\end{eqnarray}
Here $\rho_{I} = T + i\pi \alpha_{1}$ (and $\bar{\rho}_{I}$) is
the interaction point on the $\rho$-plane
and $\sigma^{(3)}_{I}$ denotes the value of the $\sigma_{3}$
coordinate of the interaction point of string $3$.
We set $\sigma^{(3)}_{I}=\pi \alpha_{1}/\alpha_{3}$ so that
\begin{equation}
\beta_{3} = 0~.
\end{equation}

The topology of the string diagram Fig.\ref{fig:annulus1}
is an annulus with a puncture corresponding to string $3$.
Therefore the $\rho$-plane
can be mapped to a rectangle with a puncture
on the complex $\nu$-plane~(Fig.\ref{fig:rectangle}).
We take this rectangle to be the region 
defined by $-\frac{1}{2} \leq \mathrm{Re}\, \nu \leq 0$
and $-\frac{\tau}{2}\leq \mathrm{Im}\,\nu \leq \frac{\tau}{2}$.
Here $\tau$ $(\tau \in i \mathbb{R})$ is the moduli parameter
and the identification $\nu \cong \nu + \tau$ should be made.
These two surfaces are related by the Mandelstam
mapping\footnote{The Mandelstam mapping (\ref{eq:rectangle})
is essentially the same as the one in \cite{Kishimoto:2004km}.
The rectangle on the $\nu$-plane introduced here
is the dual annulus of the rectangle on the $u$-plane
considered in \cite{Kishimoto:2004km}.
These are related by $\nu = \frac{u}{\tilde{\tau}}$, where
$\tilde{\tau}=-\frac{1}{\tau}$.
See also \cite{Asakawa:1999nc}\cite{Kishimoto:2006gd}
\cite{Giddings:1986rf}.}
\begin{equation}
\rho (\nu)
= \alpha \ln \frac{\vartheta_{1} (\nu + V_{3} | \tau)}
                  {\vartheta_{1} (\nu - V_{3}|\tau)}~,
\label{eq:rectangle}
\end{equation}
where $\alpha =\alpha_{1} + \alpha_{2} = -\alpha_{3} >0$,
$V_{3}=-\frac{\alpha_{1}}{2\alpha}$ and $\vartheta_{i}(\nu|\tau)$
$(i=1,\ldots,4)$ are the theta functions.
The point $\nu = V_{3}$ is the puncture corresponding to
the origin $w_{3}=0$ of the unit disk $|w_{3}|\leq 1$ of
 string~$3$.
We may parametrize the interaction points $\nu^{-}_{I}$ and $\nu^{+}_{I}$
on the $\nu$-plane corresponding to $\rho_{I}$ and $\bar{\rho}_{I}$
on the $\rho$-plane as $\nu^{\pm}_{I} = -y \mp \frac{\tau}{2}$ with
$y \in \mathbb{R}$, $0\leq y \leq \frac{1}{2}$.
These are determined by $\frac{d\rho}{d\nu} (\nu_{I}^{\pm}) = 0$.
This yields
\begin{equation}
g_{4} \left.\left(\frac{\alpha_{1}}{2\alpha} + y \,\right|\tau\right)
+g_{4}\left.\left(\frac{\alpha_{1}}{2\alpha} - y \,\right|\tau\right)
=0~,
\label{eq:interactionpt-y-annulus}
\end{equation}
where $g_{i}(\nu|\tau) = \partial_{\nu} \ln \vartheta_{i}(\nu|\tau)$.
The relation $\mathrm{Re}\, \rho(\nu_{I}^{\pm}) = T$ leads to
\begin{equation}
T= \alpha \ln
   \frac{\vartheta_{4}\left.\left(\frac{\alpha_{1}}{2\alpha} + y 
                      \,\right|\tau\right)}
         {\vartheta_{4} \left.\left( \frac{\alpha_{1}}{2\alpha} -y
                      \,\right| \tau \right)}~.
\label{eq:annulus-T}
\end{equation}

It follows from eqs.(\ref{eq:interactionpt-y-annulus})
and (\ref{eq:annulus-T}) that in the small $T$ limit,
$T = \epsilon \ll 1$, the parameters $\tau$ and $y$ behave as
follows~\cite{Kishimoto:2004km}\cite{Kishimoto:2006gd}:
\begin{equation}
q^{\frac{1}{2}} \equiv e^{i\pi \tau} \simeq
 \frac{\epsilon}{4\alpha \sin \left(\pi\frac{\alpha_{1}}{\alpha}\right)}
 + \mathcal{O}(\epsilon^{3})~,
\quad
y \simeq \frac{1}{4}
   -\frac{1}{\pi} \cos \left(\pi\frac{\alpha_{1}}{\alpha}\right)
    q^{\frac{1}{2}}
    + \mathcal{O} (q^{\frac{3}{2}} )~.
\label{eq:smallT-annulus1}
\end{equation}
Therefore we find that in this limit the moduli parameter $-i \tau$
becomes infinity.
For later use, we consider
the behavior of $\tau$ and $y$ in the $T\rightarrow \infty$ limit
as well. In this limit, the moduli parameter $\tau$ tends to $0$.
In fact, we have
\begin{equation}
T \sim \frac{\alpha_{1}\alpha_{2} \pi}{\alpha} \frac{i}{\tau}
        + \hat{\tau}_{0}~,
\quad
y \sim \frac{\alpha_{1}}{2\alpha}
 + \frac{i}{2\pi} \tau \ln \frac{\alpha_{1}}{\alpha_{2}}~.
\label{eq:largeT-annulus}
\end{equation}

\subsubsection*{Neumann coefficients}
The part $-\pi \alpha_{1} \leq \mathrm{Im}\,\rho \leq \pi \alpha_{1}$
of the boundary $\mathrm{Re}\,\rho = 0$ of the $\rho$-plane
where the $\rho$-plane is attached to $|B_{0}\rangle_{1}$
corresponds to the side $\mathrm{Re}\,\nu=-\frac{1}{2}$
of the rectangle on the $\nu$-plane.
The remaining part of the boundary of the $\rho$-plane
where the $\rho$-plane is attached to $|B_{0}\rangle_{2}$
corresponds to the the other side $\mathrm{Re}\,\nu=0$ of the rectangle
on the $\nu$-plane.
Therefore, on the $\nu$-plane
the worldsheet variables $X^{N}(\nu,\bar{\nu})$
satisfy the Neumann and the Dirichlet boundary conditions
according to eq.(\ref{eq:boundarycond}) on the two sides,
$\mathrm{Re}\,\nu=-\frac{1}{2}$ and $\mathrm{Re}\, \nu =0$,
of the rectangle and the periodic boundary
condition $X^{N}(\nu+\tau,\bar{\nu}-\tau) = X^{N}(\nu,\bar{\nu})$
along the imaginary axis. 
It follows that the two-point functions
of $X^{N}(\nu,\bar{\nu})$ on the $\nu$-plane become
\begin{eqnarray}
\lefteqn{G^{NM}_{\mathrm{rectan.}}(\nu,\bar{\nu};\nu',\bar{\nu}')
=\langle X^{N}(\nu,\bar{\nu}) X^{M}(\nu',\bar{\nu}')
 \rangle
} \nonumber\\
&& \quad = {}-\eta^{NM} \ln \vartheta_{1}
                     \left.\left(\nu-\nu'\, \right|\tau\right)
     -\eta^{NM} \ln \vartheta_{1}
                     \left.\left(\bar{\nu}-\bar{\nu}'\,\right|\tau\right)
\nonumber\\
&& \qquad   {}-D^{NM} \ln \vartheta_{1}
                   \left.\left( \nu + \bar{\nu}'\,\right|\tau\right)
     -D^{NM} \ln \vartheta_{1}
                   \left.\left( \bar{\nu} + \nu' \, \right|\tau\right)
   {} + f^{NM}(\nu,\bar{\nu};\nu',\bar{\nu}')~,
\label{eq:G-rectangle}
\end{eqnarray}
where $f^{NM}(\nu,\bar{\nu};\nu',\bar{\nu}')$ are the terms
necessary for the periodicity of the two-point functions
$G^{NM}_{\mathrm{rectan.}}(\nu,\bar{\nu};\nu',\bar{\nu}')$
along the imaginary axis of the $\nu$-plane, defined as
\begin{equation}
f^{\mu\lambda}(\nu,\bar{\nu};\nu',\bar{\nu}') = -\eta^{\mu\lambda}
  \frac{\pi i}{\tau} \left( \nu -\nu' -\bar{\nu}+\bar{\nu}'\right)^{2}
\end{equation}
for $\mu,\lambda \in \mathrm{N}$, and $0$ otherwise.
In eq.(\ref{eq:G-rectangle}), we have used the relation
\begin{equation}
\overline{\vartheta_{1}(\nu|\tau)} = \vartheta_{1}(\bar{\nu}|\tau)
\end{equation}
for $\tau \in i \mathbb{R}$.

The vertex $\langle V^{0}_{1}(3);T|$ can be expressed as
\begin{equation}
\langle V_{1}^{0}(3);T|
 = 2 \delta (\alpha_{1} + \alpha_{2} + \alpha_{3} ) 
  (2\pi)^{p+1} \delta^{p+1}_{\mathrm{N}} (p_{3})
 \mathcal{K}_{1}(3;T)
  \langle V^{0}_{1,\mathrm{LPP}}(3);T|~,
\end{equation}
where
$\langle V^{0}_{1,\mathrm{LPP}}(3);T|$ is the LPP vertex,
which is of the form
\begin{eqnarray}
\lefteqn{
\langle V^{0}_{1,\mathrm{LPP}}(3);T|
} \nonumber\\
 && =
  {}_{3} \langle 0| \exp \left[ \sum_{n,m=0}^{\infty}
     \Bigl\{ \frac{1}{2} \left(
    \bar{N}^{hh}_{nm,NM}\,
           \alpha^{N(3)}_{n} \, \alpha^{M(3)}_{m}
   + \bar{N}^{aa}_{nm,NM} \,
           \tilde{\alpha}^{N(3)}_{n} \tilde{\alpha}^{M(3)}_{m}
 \right. \right.
 \nonumber\\
 && \hspace{9.5em} \left. \left. + 
   \bar{N}^{ha}_{nm,NM} \, 
        \alpha^{N(3)}_{n} \tilde{\alpha}^{M(3)}_{m}
        +\bar{N}^{ah}_{nm,NM} \,
         \tilde{\alpha}^{N(3)}_{n} \alpha^{M(3)}_{m}
         \right) 
   \Bigr\}
   \right],
\end{eqnarray}
and the remaining factor $\mathcal{K}_{1}(3;T)$ is
independent of the non-zero oscillation modes.

The Neumann coefficients $\bar{N}^{hh}_{nm,NM},~\bar{N}^{aa}_{nm,NM},
~\bar{N}^{ha}_{nm,NM},~\bar{N}^{ah}_{nm,NM}$ are
determined by the equation
\begin{eqnarray}
&&
\int d'3
\langle V^{0}_{1,\mathrm{LPP}}(3);T| X^{N(3)}(w_{3},\bar{w}_{3})
  X^{M(3)}(w'_{3},\bar{w}'_{3}) |0\rangle_{3} (2\pi)^{26}\delta^{26}(p_{3})
  i\bar{\pi}^{(3)}_{0}\pi^{(3)}_{0}
\nonumber\\
&& \qquad\qquad 
= G^{NM}_{\mathrm{rectan.}}(\nu_{3},\bar{\nu}_{3};\nu'_{3},\bar{\nu}'_{3})~,
\end{eqnarray}
where $\nu_{3}$ and $\nu'_{3}$ are the points on the $\nu$-plane
corresponding to the points $w_{3}$ and $w'_{3}$
on the unit disk of string $3$ respectively.
We obtain
\begin{eqnarray}
& &
\bar{N}^{hh,NM}_{nm}
=
\left(
\bar{N}^{aa,NM}_{nm}
\right)^\ast
=
\frac{-1}{nm}
\oint_{V_3}\frac{d\nu}{2\pi i}
\oint_{V_3}\frac{d\nu'}{2\pi i}
e^{-n\zeta_3 (\nu )-m\zeta_3' (\nu' )}
\partial_\nu \partial_{\nu'}
G^{NM}_{\mathrm{rectan.}}(\nu ,\bar{\nu};\nu',\bar{\nu}')
~,
\nonumber
\\
& &
\bar{N}^{ha,NM}_{nm}
=
\left(
\bar{N}^{ah,NM}_{nm}
\right)^\ast
=
\frac{-1}{nm}
\oint_{V_3}\frac{d\nu}{2\pi i}
\oint_{V_3}\frac{d\bar{\nu}'}{2\pi i}
e^{-n\zeta_3 (\nu )-m\bar{\zeta}_3' (\bar{\nu}' )}
\partial_\nu \partial_{\bar{\nu}'}
G^{NM}_{\mathrm{rectan.}}(\nu ,\bar{\nu};\nu',\bar{\nu}')~,
\nonumber
\\
& &
\frac{1}{2}\left(
 \bar{N}^{hh,NM}_{n0}
 +
 \bar{N}^{ha,NM}_{n0} \right)
=
\frac{1}{2}
\left(
 \bar{N}^{aa,NM}_{n0}
  +
 \bar{N}^{ah,NM}_{n0}
\right)^\ast
\nonumber\\
&& \hspace{10em} 
  ={} - \frac{1}{2n}
\oint_{V_3}\frac{d\nu}{2\pi i}
e^{-n\zeta_3 (\nu )}
\partial_\nu 
G^{NM}_{\mathrm{rectan.}}(\nu ,\bar{\nu};V_3,V_3)
~,
\nonumber
\\
& &
\frac{1}{4} \left(
 \bar{N}^{hh,NM}_{00}
  +
 \bar{N}^{ha,NM}_{00}
  +
 \bar{N}^{ah,NM}_{00}
  +
 \bar{N}^{aa,NM}_{00}
\right)
\nonumber\\
&& \hspace{10em} =
 {} -\frac{1}{4}
    \lim_{\nu \rightarrow V_3}
      \left(
         G^{NM}_{\mathrm{rectan.}}(\nu ,\bar{\nu};V_3,V_3)
         +
         \zeta_3(\nu )+\bar{\zeta}_3(\bar{\nu})
      \right)~.
\end{eqnarray}

\subsubsection*{$\mathcal{K}_{1}(3;T)$}
The prefactor $\mathcal{K}_{1}(3;T)$ can be determined through the
method in \cite{Mandelstam:1985ww} again. 
We excise small semi-circles around
the interaction points $\nu=  \nu_{I}^{\pm}$ and a small circle
around the puncture $\nu=V_{3}$ on the $\nu$-plane.
This time, besides the contributions from these holes, 
we should include the moduli dependence of the partition function, 
and we find 
\begin{equation}
\mathcal{K}_{1}(3;T)
\propto
\left(
|\alpha |^{-2}
 \left| \frac{d w_3}{d \nu}
\left(
V_3
\right)
\right|^2
\right)
|c_{I}|^{-1}
|\tau |^{-\frac{p+1}{2}}
\eta (\tau )^{-24}
,
\end{equation}
where $\eta(\tau)$ is the Dedekind eta function and
$c_{I}$ is defined by
\begin{equation}
c_{I} = \frac{d^{2}\rho}{d\nu^{2}} (\nu_{I}^{-})~.
\end{equation}
Thus we obtain 
\begin{equation}
\mathcal{K}_{1}(3;T) = \mathcal{K}'_{0}
 \frac{(2\pi)^{2} e^{\frac{2T}{\alpha}}}
      {(-i\tau )^{\frac{p+1}{2}} \eta (\tau)^{18} \alpha^{2} c_{I}
        \vartheta_{1} \!\!\left.\left(\frac{\alpha_{1}}{\alpha}\,
                      \right|\tau\right)^{2}}~,
\end{equation}
where $\mathcal{K}'_{0}$ is a numerical factor which cannot be determined
by this method.
The factor $\mathcal{K}'_{0}$ can be fixed by
comparing the behaviors in the $T\rightarrow \infty$ limit
of the left and right hand sides of the following equation,
\begin{eqnarray}
&&
\int d'1d'2d'3
 \langle V_{3}^{0}(1,2,3)|B_{0}\rangle_{1}^{T}
 |B_{0}\rangle^{T}_{2}
 |0\rangle_{3} (2\pi)^{26} \delta^{26}(p_{3})
                i\bar{\pi}^{(3)}_{0} \pi^{(3)}_{0}
  \nonumber\\
&& \qquad = \int d'3 \;
     2 \delta (\alpha_{1}+\alpha_{2}+\alpha_{3})
     (2\pi)^{p+1}\delta^{p+1}_{\mathrm{N}} (p_{3})
     \mathcal{K}_{1}(3;T) 
\nonumber\\
&& \hspace{6em}
      \times
      \langle V^{0}_{1,\mathrm{LPP}} (3);T|0\rangle_{3}
       (2\pi)^{26} \delta^{26}(p_{3})
                i\bar{\pi}^{(3)}_{0} \pi^{(3)}_{0}~.
\end{eqnarray}
By making use of eq.(\ref{eq:largeT-annulus}),
we find that
\begin{equation}
\mathcal{K}'_{0}
 =  \frac{(2\pi )^{p+1}}{(2\pi)^{25}}~.
\end{equation}

\subsubsection*{$C(\rho_I)$}
Let us consider the effect of the insertion of the
ghost field $C$ at the interaction point.
In the same way as eq.(\ref{eq:Cint-tree}), this can be
described by
\begin{eqnarray}
\langle V_{1,\mathrm{LPP}}(3);T| &\equiv&
 \langle V^{0}_{1,\mathrm{LPP}}(3);T|C(\rho_{I})
 \nonumber\\
&=& \langle V^{0}_{1,\mathrm{LPP}}(3);T|
  \sum_{n=0}^{\infty} \left( 
    M_{\mathrm{rectan.} n}^{\ \ \ \ \ \ h} \,i\gamma^{(3)}_{n}
    + M_{\mathrm{rectan.} n}^{\ \ \ \ \ \ a}
       \, i\tilde{\gamma}^{(3)}_{n}
    \right)~.
\end{eqnarray}
The coefficients $M_{\mathrm{rectan.} n}^{\ \ \ \ \ \ h}$
and $M_{\mathrm{rectan.} n}^{\ \ \ \ \ \ a}$
can be determined through the LPP prescription
by requiring that
\begin{eqnarray}
\lefteqn{
\int d'3
\langle V_{1,\mathrm{LPP}}(3);T| \bar{C}^{(3)}(w_{3},\bar{w}_{3})
 |0\rangle_{3} (2\pi)^{26} \delta^{26}(p_{3})
  i\bar{\pi}^{(3)}_{0} \pi^{(3)}_{0}
}
\nonumber\\
&&
= G^{C\bar{C}}_{\mathrm{rectan.}}(\nu_{I}^{-},\bar{\nu}_{I}^{-};
   \nu_{3},\bar{\nu}_{3})
 \nonumber\\
&& 
  = i \Bigl[ \ln \vartheta_{1}(\nu_{I}^{-}-\nu_{3}|\tau) 
             + \ln \vartheta_{1}(\bar{\nu}^{-}_{I} -\bar{\nu}_{3}|\tau)
             - \ln \vartheta_{1}(\nu_{I}^{-} + \bar{\nu}_{3}|\tau)
             - \ln \vartheta_{1}(\bar{\nu}^{-}_{I} + \nu_{3}|\tau)
       \Bigr]~.~~~
\end{eqnarray}
It follows that the coefficient of the zero mode
$\gamma^{(3)}_{0}=\tilde{\gamma}^{(3)}_{0}=\pi^{(3)}_{0}$ is
\begin{eqnarray}
\lefteqn{
M_{\mathrm{rectan.} 0}^{\ \ \ \ \ \ h}
 + M_{\mathrm{rectan.} 0}^{\ \ \ \ \ \ a}
} \nonumber\\
&& 
= \ln \vartheta_{1}(\nu_{I}^{-}-V_{3}|\tau) 
             + \ln \vartheta_{1}(\bar{\nu}^{-}_{I} -V_{3}|\tau)
             - \ln \vartheta_{1}(\nu_{I}^{-} + V_{3} |\tau)
             - \ln \vartheta_{1}(\bar{\nu}^{-}_{I} + V_{3} |\tau)
\nonumber\\
&&
= 2 \ln \frac{\vartheta_{4}\left.\left(\frac{\alpha_{1}}{2\alpha} -y
              \,\right|\tau \right)}
              {\vartheta_{4}\left.\left(\frac{\alpha_{1}}{2\alpha} +y
              \,\right|\tau \right)}~,
\end{eqnarray}
and those of the non-zero modes are
\begin{equation}
M^{\ \ \ \ \ \ h}_{\mathrm{rectan.} n}
 = \left( M^{\ \ \ \ \ \ a}_{\mathrm{rectan.} n} \right)^{\ast}
 = -\frac{1}{n} \oint_{\nu=V_{3}} \frac{d\nu}{2\pi i}
    \Bigl[ g_{1} ( \nu^{-}_{I} -\nu | \tau )
           + g_{1} (\bar{\nu}^{-}_{I} + \nu | \tau ) \Bigr]
\end{equation}
for $n \geq 1$.

\subsubsection*{$\langle V_{1}(3);\epsilon |$}
Collecting all the results obtained in the above, we eventually get
the complete expression of the vertex $\langle V_{1}(3);T|$.
Let us take $T= \epsilon \to 0$ limit.
Again it is intuitively obvious and straightforward to show that
\begin{equation}
(2\pi)^{p+1} \delta^{p+1}_{\mathrm{N}} (p_{3})
\langle V^{0}_{1,\mathrm{LPP}}(3);\epsilon |
\sim
{}^{\epsilon}_{3} \langle B_{0}|~,
\end{equation}
in the leading order.
It follows that only the $\pi_0^{(3)}$ from 
$C(\rho_I)$ survives the level matching projection and thus
\begin{equation}
(2\pi)^{p+1} \delta^{p+1}_{\mathrm{N}} (p_{3})
\langle V_{1,\mathrm{LPP}}(3);\epsilon |
\mathcal{P}_{3}
\sim
{}^{\epsilon}_{3} \langle B_{0}|
\frac{2i}{\alpha_{3}}\pi^{(3)}_{0}
\mathcal{P}_{3}~.
\label{eq:CrhoItorus}
\end{equation}
Evaluating $\mathcal{K}_1(3,\epsilon )$, we eventually get
\begin{equation}
\langle V_{1}(3);\epsilon |
 \sim
  - 2\delta (\alpha_{1}+\alpha_{2}+\alpha_{3})
      \frac{(4\pi^{3})^{\frac{p+1}{2}} }{(2\pi)^{25}}
      \frac{4}{\epsilon^{2}(-\ln \epsilon)^{\frac{p+1}{2}}}
      {}^{\epsilon}_{3} \langle B_{0}|
      \frac{2i}{\alpha_{3}}\pi^{(3)}_{0} \mathcal{P}_{3}
      ~.
\label{eq:v1(3)}
\end{equation}

One can treat the case $\alpha_1 ,\alpha_2 <0$ in the same way and 
show that eq.(\ref{eq:v1(3)}) also holds in this case. 

We would like to comment on the calculations in \cite{Baba:2006rs}.
In \cite{Baba:2006rs}, essentially a quantity such as 
\begin{equation}
\int d'1 d'2 d'3 \,
\langle V_3(1,2,3)|
   \bar{\pi}_0^{(r)}
   |B_0\rangle_1^\epsilon |B_0\rangle_2^\epsilon 
   |B_0\rangle_3^\epsilon~,
\label{eq:bbb}
\end{equation}
is calculated to express the BRST transformation of the solitonic operators 
in terms of the string fields expanded by the normalized boundary state. 
Here let us consider the situation $\alpha_1\alpha_2>0$. 
This quantity can be calculated using either 
$\langle V_2(1,2);\epsilon |$ or $\langle V_1(3);\epsilon |$ by 
taking overlaps with $|B_0\rangle^\epsilon $'s\footnote{
Notice that one should not use eqs.(\ref{eq:v2(12)})(\ref{eq:v1(3)})
to calculate eq.(\ref{eq:bbb}). 
Eqs.(\ref{eq:v2(12)})(\ref{eq:v1(3)}) hold in the leading order in $\epsilon$ 
and we have $\mathcal{O}\left(\frac{-1}{\ln \epsilon}\right)$ corrections. 
}.
Here let us devote our attention to the effect of the insertion
of $C(\rho_I)$ in $\langle V_3(1,2,3)|$.  From eq.(\ref{eq:CrhoItree})
we can see that with $\epsilon$ small, 
\begin{equation}
C(\rho_I)\sim 
\left(
\frac{i}{\alpha_1}\pi_0^{(1)}+ \frac{i}{\alpha_{2}} \pi^{(2)}_{0} 
\right)~.
\end{equation}
On the other hand, from eq.(\ref{eq:CrhoItorus}) one can see that 
\begin{equation}
C(\rho_I)\sim 
\frac{2i}{\alpha_{3}}\pi^{(3)}_{0}~.
\label{eq:pi3}
\end{equation}
Therefore the effect of inserting $C(\rho_I)$ is a bit asymmetric 
among 1st, 2nd and 3rd strings depending on the sign of $\alpha_r$. 
These effects were overlooked in \cite{Baba:2006rs}, 
and we took $C(\rho_I)\sim 
\frac{i}{\alpha_{3}}\pi^{(3)}_{0}$
instead of eq.(\ref{eq:pi3}), to calculate eq.(\ref{eq:bbb}).
With these effects taken into account, 
the results in \cite{Baba:2006rs} are consistent with the ones here.

\newpage

\end{document}